# Dimensional scaffolding of electromagnetism using geometric algebra

**Xabier Prado Orbán[1] and Jorge Mira[2]**

[1] Departamento de Didácticas Aplicadas, Universidade de Santiago de Compostela, E-15782 Santiago de Compostela, Spain

[2] Departamento de Física Aplicada – área de electromagnetismo, Universidade de Santiago de Compostela, E-15782 Santiago de Compostela, Spain

E-mail: jorge.mira@usc.es

**Abstract**

Using geometric algebra and calculus to express the laws of electromagnetism we are able to present magnitudes and relations in a gradual way, escalating the number of dimensions. In the one-dimensional case, charge and current densities, the electric field E and the scalar and vector potentials get a geometric interpretation in spacetime diagrams. The geometric vector derivative applied to these magnitudes yields simple expressions leading to concepts like divergence, displacement current, continuity and gauge or retarded time, with a clear geometric meaning. As the geometric vector derivative is invertible, we introduce simple Green´s functions and, with this, it is possible to obtain retarded Liénard-Wiechert potentials propagating naturally at the speed of light. In two dimensions, these magnitudes become more complex, and a magnetic field B appears as a pseudoscalar which was absent in the one-dimensional world. The laws of induction reflect the relations between E and B, and it is possible to arrive to the concepts of capacitor, electric circuit and Poynting vector, explaining the flow of energy. The solutions to the wave equations in this two-dimensional scenario uncover now the propagation of physical effects at the speed of light. This anticipates the same results in the real three-dimensional world, but endowed in this case with a nature which is totally absent in one or three dimensions. Electromagnetic waves propagating exclusively at the speed of light in vacuum can thus be viewed as a consequence of living in a world with an odd number of spatial dimensions. Finally, in the real three-dimensional world the same set of simple multivector differential expressions encode the fundamental laws and concepts of electromagnetism.

## 1. Introduction

The concept of a geometric algebra was first established by Grassmann [1, 2] to encode the main elements of a geometric space. Clifford [3, 4] later added the quaternions developed by Hamilton [5, 6, 7], which encoded the operations on these same spaces [8], together with the main concept of a geometric product between these elements. Greatly overshadowed [9] by the vector algebra developed at the time by Gibbs [10] and Heaviside [11], which was thoroughly employed to express and to manipulate Maxwell´s equations for electromagnetism, geometric algebra has nonetheless survived and it has been revived by authors like Hestenes [12, 13] -which has shown that the geometric product and its algebraic properties can be efficiently applied to differential operators - as a new language for mechanics, electromagnetism and modern physics .

Readers who are not familiar with the geometric algebra of spacetime should be careful to avoid confusion with the usual 3-D geometric algebra. Appendix A presents the main concepts and relations such as multivectors, geometric product, spacetime directions, indexes and their rules, vector derivative, space-time split, relative vectors, spatial directions, and cross product.

In this framework, the set of Maxwell´s equations can be unified in the simple expression $\nabla F = J$ (see section A1.3 of Appendix A for the definition of F, J and $\nabla$), which is the electromagnetic field equation (or simply, field equation) in geometric algebra. With the help of a space-time split, this formula can be translated to relative vectors, getting in this way the four Maxwell equations in the usual vector formulation. Natural units (like the Heaviside-Lorentz ones) where $\mu_0 = 1$ and $\varepsilon_0 = 1$ are being used consistently and, consequently, the speed of light has the value c=1.

As explained in Appendix A, we will use the symbols $\gamma_i$ and $\sigma_i$ to represent directions in spacetime and in the associated space. In the case of multivector directions, the subindexes follow the convention $\gamma_{ij...} = \gamma_i \gamma_j ...$ and $\sigma_{ij...} = \sigma_i \sigma_j ...$ (1.1)

This is a conveniently compact way to express the base elements of the spacetime algebras throughout this paper.

The product of the base vectors is called the pseudoscalar, which is a n-vector encoding the directionality of the whole space.

The symbol i identifies the euclidean pseudoscalar ($i_n$ in our notation), in order to keep the expressions valid for dimensions other than 3.

In order to apply a space-time split to the field bivector F we consider the following expressions (see Appendix A.2 for more details):

$F = \mathbf{E} + i\mathbf{B}$ (1.2)

$\gamma_0 F \gamma_0 = \mathbf{E} - i\mathbf{B}$ (1.3)

where $\mathbf{E} = \frac{1}{2}(F - \gamma_0 F \gamma_0)$ anticommutes with $\gamma_0$, and

$i\mathbf{B} = \frac{1}{2}(F + \gamma_0 F \gamma_0)$ commutes with $\gamma_0$.

**E** is the electric field corresponding to the temporal component of the bivector F, and the product i**B** corresponds to the purely spatial part of the spacetime bivector F. We identify B with the magnetic field. In all dimensions, the electric field is a relative vector, but the magnetic field is a relative vector only in three dimensions. In the two-dimensional world the magnetic field is a scalar magnitude, and in the one-dimensional world it is totally absent.

The expression (1.2) is formally equivalent to the Riemann-Silberstein complex representation for the electromagnetic field [14, 15, 16]), although the pseudoscalar i in the geometric algebra of spacetime has a geometric meaning which is more general than the abstract imaginary square root of -1 used in the complex representations of the EM field.

The bivector F can be derived from a vector-valued potential function A by means of the formula: $\nabla A = F$ together with a gauge to eliminate ambiguities. For reasons that will be discussed later, we apply the Lorenz gauge: $\nabla \cdot A = 0$ (1.4)

The form of the field equation in geometric algebra using this vector potential is $\nabla F = \nabla(\nabla A) = \nabla^2 A = J$ (1.5)

The rest of this paper is devoted to the application of this geometric algebraic strategy to spaces of increasing dimensions, where we will use the convention n-D to refer to a space of n dimensions and (1, n) to its corresponding spacetime. The formulation for electromagnetism as provided by an unified language for Physics [17], developed by Hestenes with the names of geometric algebra [18, 19] and geometric calculus [20, 21] provides us with a mathematical tool capable of working in different dimensions, and we will use this capability in order to present the main electromagnetic concepts in a sequential way, beginning with the lowest number of dimensions (1,1) and reaching finally its highest order (1,3). This process was named by Wheeler [22] as dimensional reduction, although using a different mathematical approach.

As R. Penrose wrote [23], if one lives in spacetime of a particular sort, the classical electrical and magnetic fields develop naturally as a consequence. We will try to show how this is made possible in spacetimes of increasing dimensions.

In every step of this dimensional escalating sequence we will begin with the field equation $\nabla F = J$ (1.6)

and separate it in multivector components, identifying the resulting expressions as the equivalents of Maxwell´s laws for this dimensional world. After applying the corresponding spacetime splits, we get these laws expressed in the more conventional vector algebra formulation, allowing us to identify the relevant electromagnetic magnitudes and their properties. In every step we will rely –if not otherwise stated- in the results obtained in the previous –lower- dimension, which explains the name of dimensional scaffolding applied to the resulting dimensional increasing sequence.

The resulting concepts and relations will be illustrated in a visual way using geometric figures in purely spatial or spacetime diagrams. The sequence of diagrams, beginning with a one-dimensional (linear) 1-D space and its (1,1) spacetime, followed by a two-dimensional (plane) 2-D space and the corresponding (1,2) spacetime, will provide a gradual visualization of the main electromagnetic concepts. Otherwise elusive concepts like the displacement current, the retarded time and its corresponding Liénard-Wiechert potentials and fields, the null cones and the propagation of signals, will get a clear geometric meaning which will evolve gradually as we increase the dimensionality of the physical space. In a few cases – like the Green´s functions and their effects on the propagation of potentials and fields - a striking difference appears between dimensions, showing that the real physical world as we know it would be quite different in other dimensions.

## 2. Dimensional scaffolding

In order to present the electromagnetic magnitudes in a sequential way, we will use a strategy of dimensional scaffolding, beginning with a one-dimensional world and adding successive spatial dimensions to get finally the whole set of magnitudes.

This process can be understood in two complementary ways:

Either as a series of toy universes where the laws of electromagnetism keep the same overall form, or else as a series of special situations, where the symmetry of the arrangement allows us to keep some of the spatial dimensions out of the equations.

The discussion about the real number of dimensions of the Universe has been recently enriched with proposals that question the existence of three spatial dimensions at the very first moments of the Big Bang [24, 25]. The absence of gravitational waves in lower dimensions make the search for experimental evidence about those early universes theoretically possible. Under this perspective, our work could make an interesting contribution to the properties of EM waves in these early universes with lower dimensionality.

In each dimensional step (n) of the dimensional scaffolding process we begin with the orthonormal vector basis

$\{\gamma_0, \gamma_1, \ldots \gamma_n\}$ for the (1,n) dimensional spacetime algebra and, after space-time splitting, we use consequently the new orthonormal relative vector basis $\{\sigma_1, ..., \sigma_n\}$. Both bases are intimately related by $\sigma_n = \gamma_{n0}$ so there is no ambiguity.

The geometric product of all the vectors of a certain basis is called a pseudoscalar. Strictly speaking, both algebras have a different pseudoscalar, as explained in Appendix A, and we thus write $i_{1,n}$ for the pseudoscalar of the (1,n) spacetime algebra and $i_n$ for that of its associated n-dimensional even subalgebra of relative vectors.

In our analysis of the EM field equations the spacetime pseudoscalar $i_{1,n}$ does not play a significant role, but the associated pseudoscalar $i_n$ is key for expressions like the cross-product or the magnetic field. To keep track with established expressions, through the text we write it always as i, under the convention that in any particular dimensional step this has to be read as $i_n$. Since in the dimensional scaffolding process we don´t mix different dimensions, no unnecessary mistakes should arise from this simplification.

The square of the pseudoscalar is of great geometric significance, and it is given by

$$i_{1,n}^2 = i_n^2 = (-1)^{n(n-1)/2} \qquad (2.1)$$

$$i_{1,1}^2 = i_1^2 = 1 \qquad (2.2)$$

$$i_{1,2}^2 = i_2^2 = -1 \qquad (2.3)$$

$$i_{1,3}^2 = i_3^2 = -1 \qquad (2.4)$$

We can observe that $i^2$ has the same value for a (1,n) spacetime and its associated n-dimensional relative vector space. This means that the corresponding geometric properties are shared by both.

This is not the case, however, when comparing a n-dimensional space with a (1, n-1) dimensional spacetime, because their pseudoscalars do not always have the same square in spite of having the same overall dimensions. A clear example would be the comparison between a (1,1) pseudo-euclidean plane as in figure 2, with $i^2 = 1$, with a 2-dimensional Euclidean plane as in figure 3 (where $i^2 = -1$). These comparisons have a geometrical nature, and since they would affect physical magnitudes in worlds of different dimensions we don´t take them further into account in this paper.

The first step in this scaffolding process will be a one-dimensional space.

The pseudoscalars, in this case, are

$$i = i_1 = \sigma_1 = \gamma_{10} = -\gamma_{01} = -i_{1,1} \qquad (2.5)$$

In this toy universe, objects can move only in one line, with two possible directions, given by $\sigma_1$ and its opposite ($-\sigma_1$), as in figure 1.

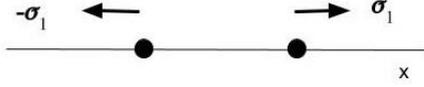

Figure 1: (1-D) one-dimensional space. Two particles are shown moving along the x-axis, one to the right ($\sigma_1$) and the other one to the left (-$\sigma_1$).

To produce an arrangement with this desired symmetry, we could think about a situation where a great cloud of uniformly distributed charges move only in one direction. This can happen, for instance, in the interior of a plasma cloud, or in a waveguide.

Adding the time dimension, this produces a (1,1) spacetime. Figure 2 shows a diagram of this one-dimensional spacetime, which is divided vertically between past and future, being the present the real world, represented by the horizontal x-axis, with a spatial direction $\gamma_1$. The vertical direction $\gamma_0$ represents the flow of time. A dashed line stands for lightlike (or null) directions. In natural units, the speed of light c =1, corresponds to those diagonal lines with unitary slope. Two curved arrows try to show the effect of a new bivector magnitude 
$$\gamma_{01} = \gamma_0\gamma_1 \qquad (2.6)$$

The fact that $\quad \gamma_{01}\,\gamma_0 = \gamma_1 \quad$ and $\quad \gamma_{01}\,\gamma_1 = \gamma_0 \qquad (2.7)$

shows that a differential operator with the spacetime bivector direction $\delta\gamma_{01}$ acting on a vector $a = a_0\gamma_0 + a_1\gamma_1$

would produce a change in the vector direction given by

$$\delta\gamma_{01}\, a = \delta\gamma_{01}\, (a_0\gamma_0 + a_1\gamma_1) = \delta(a_0\gamma_1 + a_1\gamma_0). \qquad (2.8)$$

In other words, the direction of change is symmetric to the original vector direction with respect to the diagonal. This is known as a differential boost, and the figure resulting from successive boosts (for example, in an accelerated movement) is an equilateral hyperbola as described by Minkowski [26, 27]. This is the reason to represent the bivector direction $\gamma_{01}$ as a hyperbolic arc. The observable effect of an electric field is the acceleration it induces on a test charge, which gives this symbol a clear physical meaning: it represents the direction in which a positive test charge would accelerate due to this bivector field. In contrast, when representing a purely spatial bivector in the euclidean plane, we will use a circular arc, which suggests an induced circular motion. This will be visually clear when comparing figure 2 with figure 3.

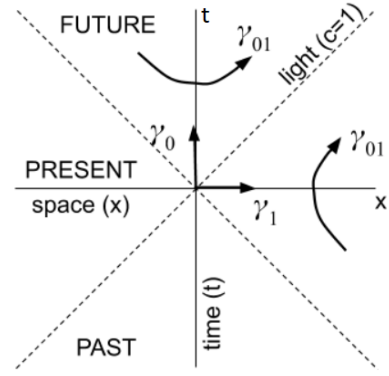

Figure 2: (1,1) bidimensional spacetime corresponding to a one-dimensional world represented by the horizontal x-axis as in figure 1. The vertical t-axis represents time increasing upwards, from the past (down) to the future (up) through the present (x-axis). In natural units, the speed of light (c = 1) corresponds to the dashed diagonal lines. Three directions are shown with arrows: a timelike direction ($\gamma_0$), a spacelike direction ($\gamma_1$), and a spacetime bivector direction ($\gamma_{01}$). The curved line for $\gamma_{01}$ over the x-axis is a hyperbolic arc which follows the path of a particle with a positive acceleration, and the one over the t-axis shows the time dilation during a spacetime rotation. Also known as boosts or Lorentz transformations, spacetime rotations have the speed of light as an asymptotic limit, as is shown by both arrows curving toward the diagonal line.

The second step in the dimensional scaffolding process is a two-dimensional space. In this toy universe, objects can move on a plane. All the possible directions are given by the linear combinations of two basis elements: ($\sigma_1$, $\sigma_2$).

Geometric algebra, as is shown in Appendix A, introduces a new kind of direction in this two-dimensional space, a bivector one, with two possible orientations, given by the pseudoscalar $i = i_2 = \sigma_{12}$ and its opposite $-i = \sigma_{21}$.

Multiplying the pseudoscalar with the spatial directions:

$$i\, \sigma_1 = \sigma_{12}\, \sigma_1 = \sigma_{121} = -\sigma_{112} = -\sigma_2 \qquad (2.9)$$

$$i\, \sigma_2 = \sigma_{12}\, \sigma_2 = \sigma_{122} = \sigma_1 \qquad (2.10)$$

Pseudoscalar magnitudes act thus as rotating operators, either clockwise ($\sigma_{21}$) or anti-clockwise ($\sigma_{12}$), like those shown in figure 3. Comparing with figure 2, we observe that there were two privileged directions (the diagonals) corresponding to the null cones or light cones. Figure 3, however, does not show any privileged or special direction. The bivector attitude in figure 3 is circular, corresponding to plane rotations, whilst the bivector attitude in figure 2 is hyperbolic, corresponding to spacetime boosts.

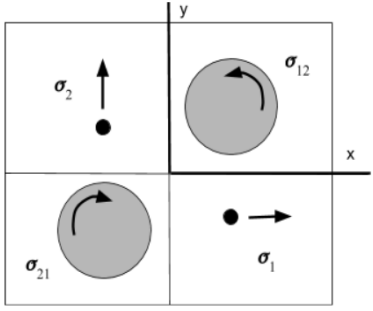

Figure 3: (2-D) two-dimensional space. A second space direction (y) has been added to the x-axis of figure 1. Two particles are shown, one moving horizontally in the x-direction ($\sigma_1$) and the other one moving vertically in the y-direction ($\sigma_2$). Two grey circles represent the unique bivector direction, with $\sigma_{12}$ and $\sigma_{21} = -\sigma_{12}$ showing its two possible orientations. The circular arrows reflect the rotating effect associated with spatial bivectors. This should be compared with the hyperbolic shape of the spacetime bivector arrows in figure 2.

This two-dimensional toy universe is equivalent to an arrangement of indefinitely long charged rods placed perpendicularly to a given plane, which would be our two-dimensional space. This can happen, for instance, in the interior of a large coil, or in the vicinity of a planar current.

Together with the time dimension, this produces a (1,2) spacetime (figure 4).

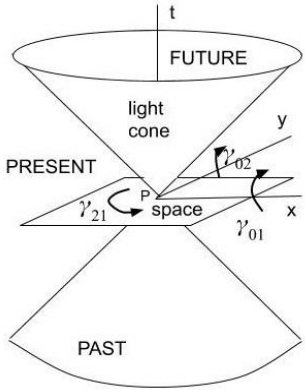

Figure 4: (1,2) three-dimensional spacetime corresponding to a bidimensional world represented by the horizontal x and y axes as in figure 3. The vertical t-axis represents time as in figure 2. The diagonal lines corresponding to the speed of light create the so-called light cone. Three bivector directions are shown with curved arrows: two hyperbolic spacetime bivectors ($\gamma_{01}$ and $\gamma_{02}$) as the one in figure 2, and a circular space bivector as in figure 3 ($\gamma_{21} = \sigma_{12}$). If we cut figure 4 vertically or horizontally, we get figures 2 and 3 respectively.

Time is the vertical axis, space is a horizontal plane with coordinates (x,y) and light cones separate two interior parts (future and past) from the exterior part with all the possible "presents" for a given spacetime point (event) P.

Two timelike bivectors ($\gamma_{01}$, $\gamma_{02}$) with the same meaning as in figure 2 are also shown, as well as the spacelike bivector $\gamma_{21}$ whose effect is equivalent to $\sigma_{12}$ due to the fact that

$$\sigma_{12} = \gamma_{10}\gamma_{20} = -\gamma_{1002} = -\gamma_{12} = \gamma_{21} \qquad (2.11)$$

A three-vector direction $\gamma_{012}$ would be the pseudoscalar in this spacetime

$$i_{1,2} = \gamma_{012} = -\gamma_{021} = -\gamma_0\sigma_{12} = -\gamma_0 i_2 = -\gamma_0 i \qquad (2.12)$$

Finally, the third step in our dimensional scaffolding corresponds to the real three-dimensional space and the (1,3) Minkowski spacetime.

The pseudoscalar, in this case, would be $i = i_3 = \sigma_{123}$ \qquad (2.13)

It is equal to the corresponding spacetime pseudoscalar too:

$$i = \sigma_1\sigma_2\sigma_3 = \gamma_{10}\gamma_{20}\gamma_{30} = \gamma_{01}\gamma_{20}\gamma_{03} = \gamma_{0123} = i_{1,3} \qquad (2.14)$$

In any of the dimensional scaffolding steps we will work initially with spacetime multivectors on the field equation, separating them in different multivector directions. Using the correspondences obtained at the space-time split, we will express later the same equations using their relative vector electric and/or magnetic components, which yields one of the Maxwell equations. In this way we assure both formal coherence due to the fact that the field equation does not change at all in this process (except for the reduction in dimensions), and at the same time we keep the physical flavour of the well-known Maxwell expressions with electric and magnetic vectors.

We will try to represent visually the main results in this process, using spacetime or purely spatial diagrams depending on the dimensional step N being considered.

N=1: The more interesting illustrations for the one-dimensional step correspond to the (1,1) spacetime diagrams because they show the effect of the non-euclidean spacetime geometry on the EM magnitudes. The corresponding one-dimensional spatial diagrams are trivially simple.

N=2: In the two-dimensional scaffolding step, both diagrams are complementary: 2-dimensional (plane) figures allow us to represent the physical situation being considered, and (1,2) (three-dimensional) diagrams are sometimes used to emphasize specific spacetime geometric structures like the null-cones. Perspective conventions are needed, which make them more cumbersome.

N=3: In the three-dimensional step, (1,3) spacetime diagrams would have four dimensions, and they cannot be represented as an image. The corresponding 3-dimensional diagrams are ubiquitous in the EM literature, so we didn´t bring them to the reader´s consideration here.

The following expressions will try to exploit the fact that geometric algebra allows to write down most of its equations in a synthetic, coordinate-free flavour. A great advantage of this approach is that it enables us to get rid of what Hestenes [28] called the "coordinate mathematical virus", meaning the widespread misbelief that vectorial calculations need to be made using coordinates all the way.

In Appendix B the same expressions are developed using coordinate expansions, in order to get into the subtleties of these equations.

## 3. One-dimensional electromagnetism

The first step in this dimensionally scaffolding sequence corresponds to a (1,1) version of electromagnetism where magnitudes like charge, current, electric field, divergence, null cones or waves, as well as some of their relations, will be introduced in a straightforward way taking advantage of simple spacetime diagrams in two dimensions. A similar approach has been already used to introduce concepts like the mass-energy equivalence or the Compton effect in a visual way [29].

The electromagnetic multivectors taking part at the field equation in this one-dimensional case are:

$$\nabla = \partial_0 \gamma_0 + \partial_1 \gamma_1$$

$$F = F_{01}\gamma_{01}$$

$$J = J_0\gamma_0 + J_1\gamma_1 \qquad (3.1)$$

The space-time split splits a vector into a scalar plus a bivector. The bivector acts like a vector in Euclidean space (it is thus a relative vector). See Appendix A.2 for details.

$$J\gamma_0 = J_0 + J_1\sigma_1 = \rho + (-\mathbf{j}) \qquad (3.2)$$

The minus sign for the current relative vector **j** is needed to assure a perfect accordance between the spacetime current and its spatial counterpart, as shown in Appendix B. It means that the spacetime vector J is spatially reflected with respect to the current relative vector **j**, and this does not imply any change in the latter.

$$\nabla\gamma_0 = \partial/\partial t + \sigma_1 \partial/\partial x = \partial/\partial t + \boldsymbol{\nabla} \qquad (3.3)$$

$$F = F_{01}\gamma_{01} = F_{01}\sigma_1 = E\sigma_1 = \mathbf{E} \qquad (3.4)$$

and there is no equivalent to a magnetic field in one dimension.

The field equation (Eq. 1.6) in this one-dimensional world would be: $\nabla F = J = J_0\gamma_0 + J_1\gamma_1$ (3.5)

which can be separated in two multivector directions

timelike vectors: $(\nabla \cdot F)_t = \partial_1\gamma_1 F_{01}\gamma_{01} = J_0\gamma_0$ (3.6)
spacelike vectors: $(\nabla \cdot F)_s = \partial_0\gamma_0 F_{01}\gamma_{01} = J_1\gamma_1$ (3.7)

### 3.1 Gauss´s law in one dimension: electric field and charge density

The first part of the one-dimensional field equation corresponds to the timelike component of the current vector
($J_0$): $\partial_1\gamma_1 F_{01}\gamma_{01} = \partial_1 F_{01}\gamma_0 = J_0\gamma_0$ (3.8)

implying $\partial_1 F_{01} = J_0$ (3.9)

Translating to spatial components: $\partial E / \partial x = \rho$ (3.10)

or, equivalently: $\boldsymbol{\nabla} \cdot \mathbf{E} = \rho$ (3.11)

This equation is known as the differential form of Gauss´s law. It states the strength of the electric field as due to the amount of charges, which act as the sources of the electric field.

#### 3.1.1 Visualization of Gauss´s law in one dimension

Figure 5 shows a set of 8 electric charges (for the sake of simplicity, we suppose that all charges are positive and equal in value) at rest in a given reference frame. The figure shows that the charge density ρ (which is inversely proportional to the separation between charges) is greater to the right than to the left. We can see also the electric field **E** as a relative vector pointing to the direction $\sigma_1$ at the right side, and to the opposite direction $-\sigma_1$ at the left side. Those relative vectors show the direction of the force which would feel a test particle with positive charge placed in these positions.

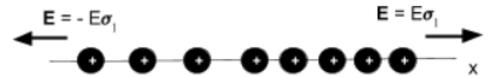

Figure 5: (1-D) space as in figure 1 with a set of particles at rest with positive charge. The distance between particles is greater to the left side of the figure. In the continuum approach, this would be observed as a charge density which increases to the right. The field E points to the right on the right side, and to the left on the left side. It is possible to identify visually the divergence of the field in the figure.

The corresponding spacetime diagram can be seen in figure 6, where the set of charges at rest is represented by parallel vertical lines. The charge density between charges is ρ = 0.

This is the reason to represent the corresponding constant field strength E as horizontal segments between charges.

We can recognize that the value of the electric field increases horizontally as if it were a counter of charges. The Gauss equation $dE/dx = \rho$ states that the charge density is equal to the slope of the function $E(x)$. In figure 6 this function is made up by successive steps (Heaviside functions). The slope of the Heaviside function is the Dirac delta function, for this reason the charge density would be in this case a sum of Dirac delta functions, having the value 0 between the charges. The grey shapes in the figure try to represent how the charge density, which is inversely proportional to the spacing between adjacent charges, would look like if the distance between individual charges were negligible (continuum approach).

In this case, where the charges are considered to be points, the E function is an increasing step function, resulting from the integration of a series of delta functions. This would be a good point to introduce these functions to undergraduate students. In the continuum approach, the steps vanish and the function $E(x)$ would be represented by a continuous line (dotted in the figure). Strictly speaking, we should use distribution theory [30, 31] to apply a differential equation to discrete charges, but we will use always the continuum approach, where the charge densities are approximated by smooth functions and the ordinary differentiation is applicable.

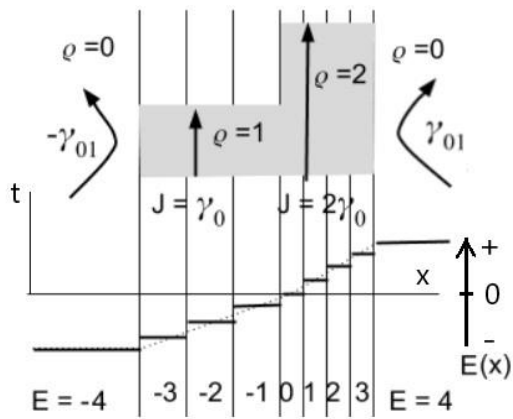

Figure 6: (1,1) bidimensional spacetime as explained in figure 2, with charges at rest as in figure 5. The fixed positions of the charges are represented by vertical lines. An additional vertical axis without geometric meaning is superimposed at the bottom, representing the strength of the electric field E as a function of the position (x). The height of the grey shapes corresponds to the charge density in the continuum approach, which will be followed throughout the rest of this paper. Explanations in full detail for this figure can be found in the text.

The current vector J is represented as an arrow in two cases (both vertically, because the current component is taken to be zero).

The strength of the field in each segment is indicated by the numbers at the bottom and the resulting graph for the field strength E in figure 6 is that of a step function. The field strength is constant (4 or -4) beyond the position of the set of charges. The way the strength of the field E measures the amount of charges is independent of the spacing of the charges. In the figure, we begin the counting at the centre of the distribution of charges (or, in other words, at the median of the charge density function), where the value of the field is E = 0. We can count in the figure four charges to its right, so the field strength at the right part of the figure will have a value of + 4. The positive sign means that the direction of the bivector E is $\gamma_{01}$. The hyperbolic arc represents an accelerated movement to the right, meaning that a positive test charge placed in this point would be repelled even farther right by the distribution of positive charges. We can count other four charges to the left of the central point. In this case, the counting yields increasingly negative values, reaching the value - 4 at the left side of the figure. The field bivector, in this case, has the direction $-\gamma_{01}$. The hyperbolic arc represents again the accelerated movement to the left a positive test charge would undergo due to the repulsion by the distribution of positive charges.

The hyperbolas representing the bivector F on both sides of figure 6 have a visually recognizable divergent attitude. This is a good starting point to explain the concept of divergence in this early step.

It is interesting to note that the Gauss equation can be interpreted also as the way in which the field E derives from the electric charges which are its sources.

In this case, where all the charges have the same sign, the electric field is proportional to the amount of charges included in the interior of a Gauss surface (in one dimension, the surface is formed by a pair of points bounding the charges).

We can even analyse how this applies to a situation with the needed symmetry, as it would be the case of having an arrangement of several charged plates. Figure 7 shows two different cases: the left case corresponds to a pair of infinite plates charged with the same sign. As in the previous figure, the field **E** diverges outside the plates, and in this case it is equally null between the plates. The right case can be used to present the concept of a capacitor, made by two plates (which in this one-dimensional world would be reduced to two points) charged with opposite signs. Gauss´s law can be applied to see that the field is only nonzero between the plates, where it points from the positive to the negative charges.

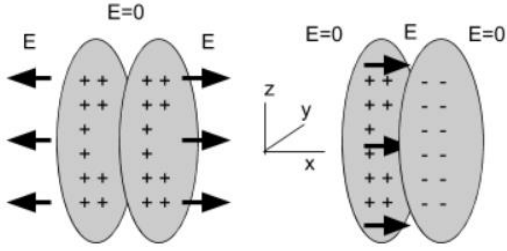

Figure 7: (3-D) space with charged plates. The common assumption that the surface of the plates has an infinite (or at least very great) extension allows us to consider only the magnitude of the electric field E in one dimension, corresponding to the x-axis.

## 3.2 Ampère´s law in one dimension: charge and displacement currents

The second part (Eq 3.7) of the one-dimensional field equation corresponds to the spacelike component of the current vector ($J_1$):

$\partial_0 \gamma_0 F_{01} \gamma_{01} = \partial_0 F_{01} \gamma_1 = J_1 \gamma_1$, implying $\partial_0 F_{01} = J_1$  (3.12)

Translating to spatial vector components: $\partial E / \partial t = -j$

Equivalently, in relative vectors: $\partial \mathbf{E}/\partial t = - \mathbf{j}$  (3.13)

We can use this result to introduce the concept of displacement current, $\mathbf{j}_D = \partial \mathbf{E}/\partial t$  (3.14)

It was proposed ad-hoc by Maxwell in order to establish a coherent set of equations and, as we have seen, it arises naturally using geometric algebra. In the one-dimensional world it reflects the effect of a time-varying electric field on the electric current.

$j_D = J_1 = - j$  (3.15)

We could view this equation also as the form of the Ampère Law in a one-dimensional world: $j_D + j = 0$  (3.16)

This equation shows us that a "charge current" j produced by the movement of charges (or, alternatively, by the movement of the observer´s reference system) is somehow compensated by a displacement current $j_D$ which is due to the time evolution of the electric field. Under this view, the displacement current would be a chargeless current, because it has no timelike component. This apparently abstract quantity, proposed by Maxwell to get a coherent set of mathematical equations, is a magnitude with real physical effects and showing interesting properties in other dimensions, as we will see.

### 3.2.1 Visualization of Ampère´s law in one dimension

In order to visualize Ampère´s law in this one-dimensional world, we can begin with an arrangement of charges at rest but viewed from a reference system which is in movement with respect to the charges. They will no longer appear at rest but they will have a "drift" velocity due to the relative movement of our reference system.

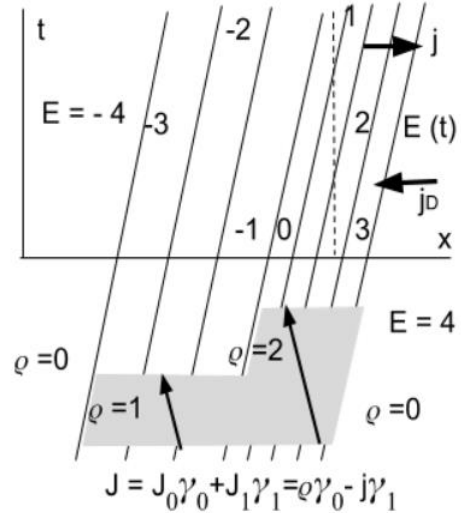

Figure 8: (1,1) bidimensional spacetime as in figure 6 after applying a boost to it. The boost is a change in the reference system due to the observer´s movement. In this case, the observer moves to the left, and as a result a drift current **j** appears to the right. The vertical lines representing the position of the charged particles at figure 6 are now accordingly inclined. The vertical dashed line helps us to identify the (negative) variation of the field E with time from 3 to 1 on this place, which gives rise to the purely spatial (horizontal) displacement current vector $\mathbf{j}_D = \partial \mathbf{E}/\partial t$. The vertical arrows which simply measured the current density in figure 6, have evolved now to the full density-current vector J. The rest of the symbols have the same meaning as in figure 6. More explanations can be found in the text.

Figure 8 represents the same situation as in figure 6, but viewed from a reference system moving to the left. The charges have now a "drift" velocity to the right. We have thus created a charge current j (to the right) which is compensated by a displacement current $j_D = \partial E/\partial t$

The figure shows that the electric field strength E is decreasing (from 3 to 1 at the right side) as time passes. The time derivative of E(t) will be negative and as a consequence the displacement current $j_D$ will point to the left.

A surprising feature in figure 8 is the fact that the charge-current vector J does not point in the direction of the particles worldlines but it is reflected horizontally, following the direction of the displacement current $j_D$ instead of the charge current j.

This reflection of the spacetime current vector with respect to the current lines is a geometric feature of spacetime that does not depend on the choice of space-time split, as shown in Appendix B.

### 3.3 Vector potential in one dimension: null directions and retarded potentials

The potential vector function in this one-dimensional world has the form $A = A_0 \gamma_0 + A_1 \gamma_1$ (3.17)

The field equation (1.5), $\nabla^2 A = J$ with this potential becomes

$$(\partial_0^2 - \partial_1^2)(A_0 \gamma_0 + A_1 \gamma_1) = J_0 \gamma_0 + J_1 \gamma_1 \quad (3.18)$$

Separating the temporal and spatial components we get a pair of very similar differential equations:

$$(\partial_0^2 - \partial_1^2) A_0 = J_0 \quad \text{and} \quad (\partial_0^2 - \partial_1^2) A_1 = J_1 \quad (3.19)$$

which are the Poisson´s equations.

The most interesting case appears when we consider a chargeless space or vacuum: $J = 0$

In this case, both differential equations reduce to the same form: $(\partial_0^2 - \partial_1^2) A = 0$ or, equivalently,

$$\partial^2 A/\partial x_0^2 - \partial^2 A/\partial x_1^2 = 0 \quad (3.20)$$

$$\partial^2 A/\partial t^2 = \partial^2 A/\partial x^2 \quad (3.21)$$

This is the Laplace equation, which is also the wave equation in one dimension, and its solutions have the form $A_R(x-t)$ or $A_L(x+t)$ corresponding to waves propagating in vacuum with speed c = 1. $A_R(x-t)$ propagates to the right, and $A_L(x+t)$ propagates to the left.

A general solution for this one-dimensional wave equation has the form

$A = A_0\gamma_0 + A_1\gamma_1$ , with

$A_0 = a_{0R}(x-t) + a_{0L}(x+t)$

$A_1 = a_{1R}(x-t) + a_{1L}(x+t)$ (3.22)

where $(a_{0R}, a_{0L}, a_{1R}, a_{1L})$ are differentiable functions in both variables (x,t). Defining a phase function φ either as

$\varphi_R = x-t \quad \text{or} \quad \varphi_L = x+t \quad (3.23)$

we have two kinds of functions with the general form a(φ), with derivatives $a'(\varphi) = da/d\varphi$.

Separating them in R and L functions

$a_R(\varphi_R) = a_R(x-t) \quad , \quad a_L(\varphi_L) = a_L(x+t) \quad (3.24)$

and applying the chain rule, we get

$\partial_0 a_R = -a'_R \quad , \quad \partial_1 a_R = a'_R$

$\partial_0 a_L = a'_L \quad , \quad \partial_1 a_L = a'_L \quad (3.25)$

The Lorenz gauge ( Eq 1.4) $\nabla \cdot A = 0$ implies $\partial_0 A_0 = \partial_1 A_1$

and, since

$\partial_0 A_0 = -a'_{0R}(\varphi_R) + a'_{0L}(\varphi_L)$

$\partial_1 A_1 = -a'_{1R}(\varphi_R) + a'_{1L}(\varphi_L) \quad (3.26)$

we arrive to

$a'_{0R} = a'_{1R} \quad \text{and} \quad a'_{0L} = a'_{1L} \quad (3.27)$

In other words, $A_0$ and $A_1$ differ only by a constant:

$A_1 = A_0 + k \quad (3.28)$

and the general expression for the one-dimensional wave potential obeying the Lorenz gauge will have the form

$A = A_0\gamma_0 + (A_0+k)\gamma_1 = A_0(\gamma_0+\gamma_1)+k\gamma_1 \quad (3.29)$

Where $A_0$ is a linear combination of $A_R(x-t)$ and $A_L(x+t)$

The electromagnetic field derived from this potential has the form

$F = \nabla A = \nabla \wedge A = (\partial_0 A_0 - \partial_1 A_0)\gamma_{01} \quad (3.30)$

After calculating

$\partial_0 A_0 = \partial_0[a_{0R}(x-t) + a_{0L}(x+t)] = -a'_{0R}+a'_{0L}$

$\partial_1 A_0 = \partial_1[a_{0R}(x-t) + a_{0L}(x+t)] = -a'_{0R}+a'_{0L} \quad (3.31)$

we arrive finally to $F = (\partial_0 A_0 - \partial_1 A_0)\gamma_{01} = 0 \quad (3.32)$

This means that every possible solution for a one-dimensional wave equation in the form of potentials will yield a null electric field. In other words, all the possible solutions do not produce any physically observable effect. As a takeaway, we could say that in a world with only one dimension there would be no electric field for the potentials in the Lorenz gauge.

The product $\nabla \cdot A$ is a scalar, and any choice of gauge different than $\nabla \cdot A = 0$ would imply the existence of a non-vanishing scalar component at the electromagnetic field. The Lorenz gauge is thus the best choice to get a compact set of equations using geometric calculus. The concept of gauge implies a degree of freedom in the choice of potential functions which render the same electromagnetic fields [32]. Our choice of the Lorenz gauge delivers potentials which behave like spacetime vectors and therefore keep an overall coherence with the relativistic framework. In classic EM, the fields are the observables, because they are directly related to the movement of test charges, so the choice of a particular gauge has only a mathematical meaning without any physically observable effect. In quantum EM, however, the potentials appear to be the real physical objects. [33, 34, 35].

It is interesting to discuss here the relation of the potentials and the null cones. The potential function A has the same form in any pair of spacetime points (events) which can be linked by a diagonal line. In other words, if the potential function has the value A in some point, it will keep this value in all the points of its null-cone.

It is useful to use the concept of retarded potential as the one that depends on the retarded time, which can be defined at this early level as $t_r = t - |x|$, which is a measure of the time lapse of the potential change from the point of view of an observer receiving its influence.

When dealing with higher dimensions, the existence of this retarded potential will be fundamental to explain the existence and properties of electromagnetic waves.

Wheeler [22, 36], using the language of p-forms, has found interesting additional properties in (1,1) dimensions which enhance what he calls the "tutorial potential" of EM as presented in this dimensionally reduced spacetime.

He assumes the absence of electric charges as an initial condition for a one-dimensional EM with electric (E) and magnetic fields (B) which generate one another under the rule of differential equations. This produces a consistent mathematical structure but at the cost of losing contact with the consideration of electromagnetism as a theory for the behaviour of electric charges.

Symmetry properties like the symplectic structure of F and the Hodge dual relationships between the E and B fields are beyond the scope of this article and should be the subject of further studies. A glimpse on this subject can be found at Appendix B.3.

The presentation of EM at this extremely simplified (1,1) level should therefore be taken with care in order to avoid unnecessary assumptions, and this is also valid for our formulation of a wave equation for the potentials. This mathematical result can be simply a consequence of our consistent choice of the Lorenz gauge from the beginning. It is only interesting from a conceptual point of view, as it allows for the presentation of the key concepts of the speed of light and retarded time at a very early stage, but, as we have shown, they do not produce any observable effect since the resulting electric field is constant. This should be stressed when presenting EM with this dimensional upgrading approach. The concept of gauge as a way to regulate redundant degrees of freedom in the Lagrangian is key for the understanding of gauge theories, which are possibly the best way to describe the interactions between charges and fields, and it could be introduced at this level to explain the difference between mathematical results and their physical meaning.

*3.3.1 Visualization of vector potentials in one dimension*

The general expression for the one-dimensional wave potential obeying the Lorenz gauge, as we have seen (Eq. 3.29), has the form

$A = A_0\gamma_0 + (A_0+k)\gamma_1 = A_0(\gamma_0+\gamma_1)+k\gamma_1$ (3.33)

Where $A_0$ is a linear combination of $A_R(\varphi_R)$ and $A_L(\varphi_L)$, being $\varphi_R$ and $\varphi_L$ the phase functions for waves propagating respectively to the right and to the left.

Figure 9 is a representation of the main magnitudes taking part in such a solution for the wave equation. It is possible to recognise that the lines of constant phase (dashed lines) correspond to worldlines with c = 1, and this is the speed of light.

We can also observe the directions $(\gamma_0+\gamma_1)$ and $(\gamma_0-\gamma_1)$ for any one-dimensional vector potential in spacetime which is a solution of the wave equation with the Lorenz gauge. These are null vectors, and we have also seen that such null vector potentials do not produce any field F, which implies that there are no physical effects associated with such potential waves in a one-dimensional spacetime.

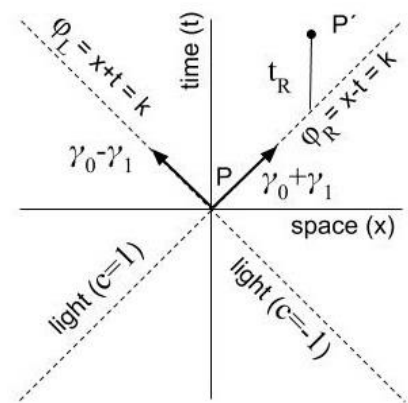

Figure 9: (1,1) spacetime as explained in figure 2, showing potential wave directions (dashed lines) and retarded time $t_R$ of P´ with respect to the source P:  $t_R = t_{P´} - r(P´- P)$.

The retarded time between the emitter P and the observer P´, $t_R$, is also shown. It reflects the time lapse (as measured by P´) from any influence coming from P. The retarded time from every point in the future light cone of P is zero.

Figure 10 shows how this can be applied to explain the redshift due to the movement of the source P from $P_1$ to $P_2$ (separated by a time delay T, which we can consider equivalent to a period for harmonic waves).

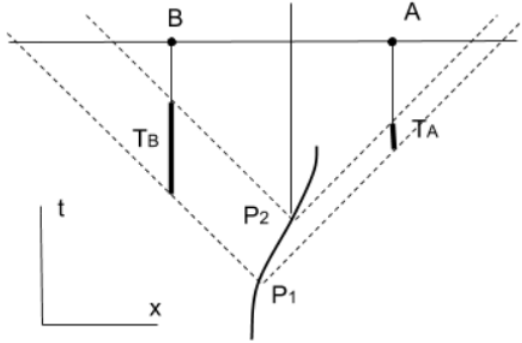

Figure 10: (1,1) spacetime as in figure 9, with the source P moving to the right from $P_1$ to $P_2$. The vertical distance between the dashed lines emitted from those two points is a measure of the period T. When the wave arrives to a receiving point A placed in front of the emitting point P this period has diminished ($T_A$) whereas the receptor B at the rear side of the source P measures a greater period ($T_B$). This is a visual explanation of the Doppler effect.

For observer A, lying in the direction of the advance of the source, the retarded time shows a delay $T_A$ while observer B, in the opposite side, measures a delay $T_B$. The redshift is clear from the figure, as $T_B > T_A$

We could view these plane figures as (1,1) sections of a general (1,3) solution, but at the cost of having to do previously all the mathematical treatment for three-dimensional EM waves and without a clear visual or intuitive meaning. The fact that we arrived to these figures with simple mathematical arguments in a one-dimensional world enables a clear understanding of an essential geometric feature of spacetime, which is the presence of privileged directions (diagonals in one dimension, null cones in greater dimensions), where the propagation of EM effects at the speed of light fits neatly.

Wheeler [22] has shown that even in this one-dimensional case it is possible to present in a visual way Green´s functions, which are the key elements to construct retarded potentials and fields from given sources.

## 4. Two-dimensional electromagnetism

We will use throughout this section the symbol i for the pseudoscalar $i = i_2 = \sigma_{12}$ (4.1)

Using Eq (2.12), $i = -\gamma_0 i_{1,2}$ (4.2)

In analogy with the alternative definition of a cross-product as explained in Appendix A.2: $\mathbf{v} \times \mathbf{w} = -\mathbf{v} \cdot (i\mathbf{w})$ (4.3)

we can define a two-dimensional vector derivative on a scalar-valued function f(x,y) as

$\nabla \times f = -\nabla(if) = \partial f/\partial_y \, \sigma_1 - \partial f/\partial_x \, \sigma_2$ (4.4)

which is coincident with the definition of a two-dimensional vector derivative -based on the notion of a perpendicular vector in two dimensions- given by McDonald [37]:

$\nabla_\perp = (\partial/\partial y \, , \, -\partial/\partial x)$ (4.5)

This is an operator with a clear geometric interpretation and a deep explanatory potential as we will see later.

The electromagnetic magnitudes taking part at the field equation in this case are:

$J = J_0 \gamma_0 + J_1 \gamma_1 + J_2 \gamma_2$

$F = F_{01} \gamma_{01} + F_{02} \gamma_{02} + F_{21} \gamma_{21}$

$\nabla = \partial_0 \gamma_0 + \partial_1 \gamma_1 + \partial_2 \gamma_2$ (4.6)

where

$\partial_0 = \partial/\partial t \, , \, \partial_1 = \partial/\partial x \, , \, \partial_2 = \partial/\partial y$ (4.7)

Figure 11 is a visualization of the vector derivative as defined for the two-dimensional space (Eq 4.4):

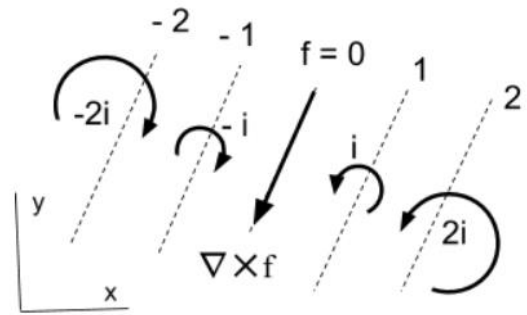

Figure 11: 2-D space as explained in figure 3 showing the effect of the vector derivative in two spatial dimensions. A bivector field **f** is represented by arcs of circles with a shape proportional to the value of f. The symbol i stands for the pseudoscalar $\sigma_{12}$ and the direction of the curved arrows shows the sign of f (positive for a counter-clockwise

direction, and negative when it is clockwise). Points with the same value for f are represented with dashed lines. The vector derivative **v**×**f** (also known as the curl of f) is an arrow which is parallel to the dashed lines and points toward the direction given by the combined effects of the bivectors on its both sides.

The two-dimensional vector derivative of a scalar field f produces a vector which is tangent to the lines of constant f and points towards the direction of the combined flux of (if) on both sides of this same line. The general situation would present curved lines, but if we assume differentiability it will be always possible to reach a scale where the curvature is negligible.

*4.1 Space-time split in two dimensions*

The two-dimensional space-time split yields the following correspondences (see Appendix A.2 for more detailed calculations):

$$J\gamma_0 = \rho + (-\mathbf{j}) \tag{4.8}$$

with $\mathbf{j} = j_x\boldsymbol{\sigma}_1 + j_y\boldsymbol{\sigma}_2$ (4.9)

$$F = \mathbf{E} + i\mathbf{B} \tag{4.10}$$

where $\mathbf{E} = E_x\boldsymbol{\sigma}_1 + E_y\boldsymbol{\sigma}_2$

and B is a scalar magnitude

$$\gamma_0 \nabla = \partial/\partial t + \nabla \tag{4.11}$$

where $\nabla = \boldsymbol{\sigma}_1 \partial/\partial x + \boldsymbol{\sigma}_2 \partial/\partial y$

is the relative vector derivative in two dimensions.

Spacetime and relative vectorial components are thus related by

$$F_{01} = E_x \;,\; F_{02} = E_y \;,\; F_{21} = B \tag{4.12}$$

$$J_0 = \rho \;,\; J_1 = -j_x \;,\; J_2 = -j_y \tag{4.13}$$

The field equation (Eq. 1.6) in this two-dimensional world would be: $\nabla F = J = J_0\gamma_0 + J_1\gamma_1 + J_2\gamma_2$ (4.14)

which can be separated in three multivector directions

timelike vectors: $(\nabla \cdot F)_t = J_0\gamma_0$ (4.15)

spacelike vectors: $(\nabla \cdot F)_s = J_1\gamma_1 + J_2\gamma_2$ (4.16)

trivectors: $\nabla \wedge F = 0$ (4.17)

*4.2 Gauss´s law in two dimensions: field divergence*

The timelike vector component (Eq. 4.15) of the two-dimensional field equation, $(\nabla \cdot F)_t = J_0\gamma_0$

implies $\partial_1 F_{01} + \partial_2 F_{02} = J_0$ (4.18)

Translating to spatial components: $\partial E_x/\partial x + \partial E_y/\partial y = \rho$

or, equivalently (as relative vectors):

$$\nabla \cdot \mathbf{E} = \text{div}(\mathbf{E}) = \rho \tag{4.19}$$

which has the same form as the usual expression for Gauss´s law. The physical meaning of this expression is, as we have seen previously for the one-dimensional case (section 3.1.1), that the sources for the electric field are the electric charges. The two-dimensional approach allows us to broaden the concept of divergence which we had already introduced in the one-dimensional world.

*4.2.1 Visualization of Gauss´s law in two dimensions*

Figure 12 represents an escalation from the (1,1)-dimensional case represented in figure 6 to a (1,2)-dimensional analogous situation. In both cases, a set of positive charges at rest are represented as vertical lines pointing upwards in spacetime. Applying Gauss´s law results in divergent bivectors on opposite sides of the charges. On the upper side of the figure the space-time split is represented as a horizontal plane with two spatial directions ($\boldsymbol{\sigma}_1$, $\boldsymbol{\sigma}_2$) and the divergence of the electric field **E** is clearly evident as outwards-pointing vectors.

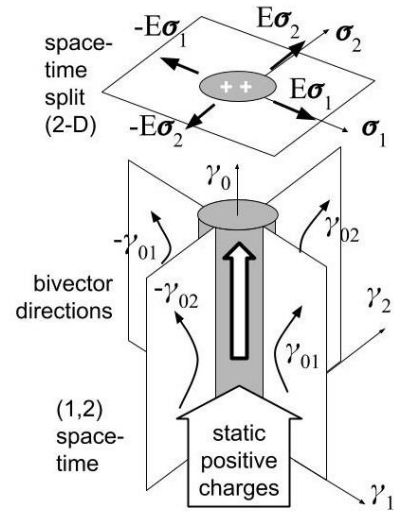

Figure 12: (1,2) spacetime as explained in figure 4 showing the field created by a group of charged particles at rest. The space directions $\gamma_1$ and $\gamma_2$ lie on the x and y axes respectively, and the time direction $\gamma_0$ points to increasing times. On the upper side can be seen the corresponding 2-D space, generated by a space-time split. The correspondence between the spacetime bivectors $\gamma_{01}$ and $\gamma_{02}$ with the space relative vectors $\boldsymbol{\sigma}_1$ and $\boldsymbol{\sigma}_2$ is clearly visible. This figure could be generated putting figure 5 over figure 6 and then rotating

the ensemble around the time axis. The divergence of the field E gets here a clearer understanding.

Taking in consideration the whole figure, it is possible to recognize that the positive charges are the source of the electric field lines, while negative charges would be its sinks. A closed line around the group of positive charges would be crossed outwards by a number of field lines which is proportional to the net amount of positive charges in its interior.

## 4.3 Ampère´s law in two dimensions: magnetic field

The spacelike component (Eq. 4.16) of the two-dimensional field equation corresponds to the spacelike components $(J_1, J_2)$ of the current vector J: $(\nabla \cdot F)_s = J_1 \gamma_1 + J_2 \gamma_2$, implying

$\partial_0 F_{01} - \partial_2 F_{21} = J_1$

$\partial_0 F_{02} + \partial_1 F_{21} = J_2$ (4.20)

Translating to relative vectors and rearranging terms we get:

$\partial E_x/\partial t + j_x = \partial B/\partial y$

$\partial E_y/\partial t + j_y = - \partial B/\partial x$ (4.21)

Recalling that B is a scalar magnitude, $iB = B\sigma_{12}$

$-\nabla(iB) = - \partial B/\partial x\ \sigma_2 + \partial B/\partial y\ \sigma_1$ (4.22)

we can write $\nabla \times B = -\nabla(iB) = \partial E/\partial t + j$ (4.23)

which can be interpreted as Ampère´s equation for a two-dimensional world.

Introducing again the displacement current $j_D = \partial E/\partial t$

Ampère´s equation can be written as

$j_D + j = -\nabla(iB)$ (4.24)

The displacement current $j_D$ appeared in the one-dimensional world (where no equivalent to a magnetic field does exist) as a compensating effect (produced by the rate of temporal change in the electric field) for a charge current. In a two-dimensional world a new component of the field (iB) must be taken into consideration. It is a pseudoscalar, and this means that we can now interpret the magnetic field as the pseudoscalar magnitude resulting from the combined effects of both currents $j_D$ and j.

This leads to an understanding of the electric field E as the timelike component of the electromagnetic field bivector F, whilst the magnetic field iB corresponds to its spacelike component.

### 4.3.1 Visualization of Ampère´s law in two dimensions

Ampère´s equation for a two-dimensional (Eqs. 4.20) world

$\partial_b F_{21} = \partial_0 F_{0a} - J_a$ (a = 1, 2) is also (Eq. 4.23)

$\nabla \times B = \partial E/\partial t + j$ (4.25)

Introducing the displacement current $j_D = \partial E/\partial t$

$-\nabla(iB) = j_D + j$ (4.26)

E is the timelike component of the electromagnetic field bivector F, and we can interpret the magnetic field iB as the pseudoscalar magnitude resulting from the combined effects of both currents $j_D$ and j (see Eq. 4.24)

We will consider first a special electrostatic situation where the electric field does not change in time: $\partial E/\partial t = 0$, and Ampère´s law can be written simply as $\nabla \times B = j$

In this case, the current relative vector j can be regarded as the source for the magnetic field. Figure 13 shows an example of the magnetic field generated by a linear current in a two-dimensional space, where we can see that the bivector field iB has opposite curling directions on both sides of the current.

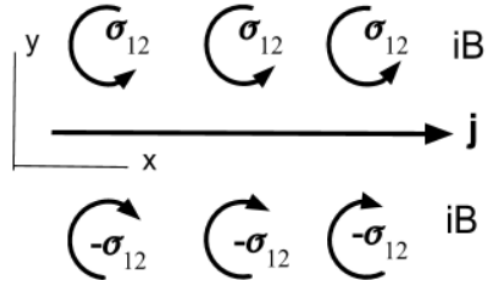

Figure 13: 2-D space as explained in figures 3 and 11, displaying a current j and the magnetic field B it gives rise to. The current can be recognized as the source for a magnetic field which rotates on both sides of the current as if they had gears in between.

Figure 14 shows a circular wire stretched by the passage of time as a cylinder in (1,2) spacetime with a steady current flowing through it -shown by several inclined particle worldliness- and producing a uniform field B in its interior. A corresponding two-dimensional space-time split is shown at the right. The same figure could be obtained by cutting a horizontal section through an indefinite vertical solenoid, which is an arrangement with the needed symmetry to be represented by a two-dimensional section showing a circular wire with a uniform current.

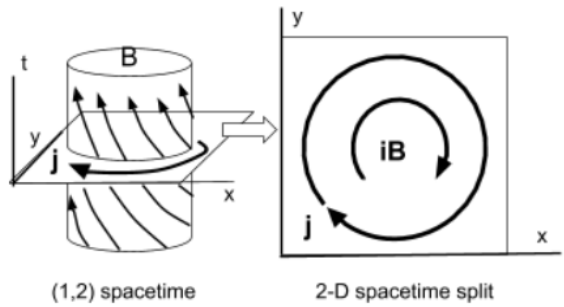

Figure 14: (1,2) spacetime as explained in figure 4, showing a two-dimensional model of a solenoid on the left side and a purely spatial section on the right side, which should be compared with figure 13 to recognize how the field B derives from the circular current j. The spacetime diagram at the left side offers a visualization of the current as a set of inclined parallel lines, much as in figure 8.

The curling direction of the field iB follows the direction of the circular current j.

It is possible to go a step further ahead in order to present the concept of an electric circuit in an alternative way, as shown in figure 15, where a battery to the left and a resistance to the right are joined by a conducting wire.

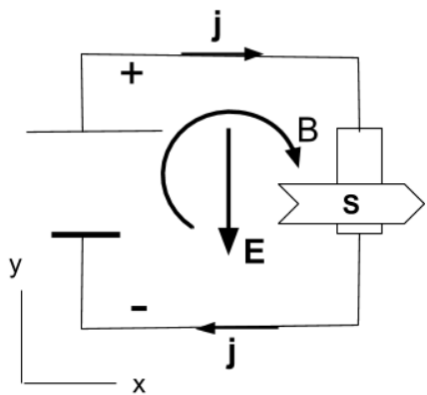

Figure 15: 2-D space as in figures 11, 13 and 14 (right side), with a model of a simple electric circuit with a battery to the left and a resistance to the right, joined by two conducting lines carrying a current with density **j**. The upper side has a positive potential and the lower side a negative one, producing an electric field **E** pointing downwards. A clockwise magnetic field B is generated as in figure 14. The product of **E** and B produces a vector **S** pointing to the right, which can be understood geometrically as the effect of B forcing the tip of **E**'s arrow to rotate clockwise. The vector **S** is known as the Poynting vector, and its physical meaning is to show the direction of the flow of energy. In this case, it is clear that the energy flows from the battery to the resistance.

We have now an additional electric field E. We represent with the letter **S** the energy flow from the battery to the resistance. It is the Poynting vector.

If the electric field is not constant, the displacement current has to be considered also as a source for the magnetic field.

This is evident in the model of a charging capacitor, as presented in figure 16.

Two parallel conducting plates (placed vertically in the figure) receive a current **j**, whose effect is to increase the charge density in the left plate and to decrease it (or to increase the negative charge density, which is equivalent) on the right plate. This produces a time-increasing electric field whose partial time derivative $\partial \mathbf{E}/\partial t$ points to the same direction as the current **j**.

The resulting displacement current $\mathbf{j}_D = \partial \mathbf{E}/\partial t$, which appeared already in section 3.2 (figure 8) as a chargeless current with a compensating effect, acts therefore again as a chargeless magnitude compensating for the absence of a material charge current in the interior of the capacitor.

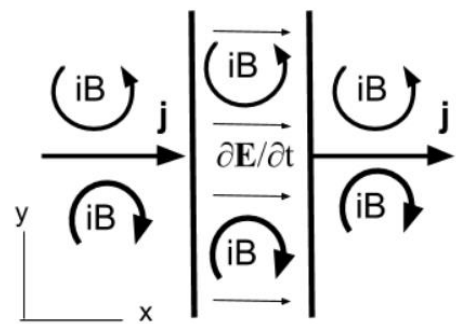

Figure 16: 2-D space as in figures 11 and 13, with a two-dimensional model of a charging capacitor. It is composed of two parallel conductors, represented by the vertical parallel lines at the centre of the figure. It can be understood as a projection of the right side of figure 7 when viewed from the z-axis. A current **j** flowing to the right is interrupted by the vacuum between both capacitors, which get increasing opposite charges as a result. The figure shows how the increasing electric field between the conductors produces a displacement current which cancels the interruption in the current **j**, and the same happens with the magnetic field B it generates in the interior of the capacitor.

If the spatial change of the electric field happens in vacuum, it generates again a pure spatial chargeless displacement current. This adds an interesting feature to the understanding of electromagnetic waves, as we will see at the end of section 4.5.1.

## 4.4 Faraday´s law in two dimensions: electromagnetic induction

The charge-current J is by definition a vector quantity, and this implies that any trivector part of the field equation must vanish identically (Eq 4.15)

In other words,  $\nabla \wedge F = 0$ (4.27)

or, equivalently,

$\partial_0 F_{21} = - \partial_1 F_{02} + \partial_2 F_{01}$ (4.28)

Translated to the corresponding relative vector magnitudes:

$\partial_t B = - \partial E_y/\partial x + \partial E_x/\partial y$ (4.29)

This equation can be further modified taking into account that

$iB = B\sigma_{12}$ (4.30)

$\nabla = \sigma_1 \partial/\partial x + \sigma_2 \partial/\partial y$ (4.31)

$E = E_x \sigma_1 + E_y \sigma_2$ (4.32)

Using again the two-dimensional vector derivative (Eq 4.4):

$\nabla \times E = -\nabla(iE) = \sigma_1 \partial E/\partial y - \sigma_2 \partial E/\partial x$ (4.33)

we get   $\partial_t B = - \partial E_y/\partial x + \partial E_x/\partial y = - \nabla \times E$ (4.34)

which is formally equivalent to Faraday´s law of induction.

### 4.4.1 Visualization of Faraday´s law in two dimensions

Figure 17 shows a spacetime diagram for a set of two circular wires, the exterior one having initially a clockwise current (**j** before) producing a negative field (B-). The current changes then to a counter-clockwise direction (**j** after) which produces a positive field (B+).

The time derivative of B (passing from B- to B+) will be thus positive.

Faraday´s equation (Eq 4.34)

$\partial_t B = - \partial E_y/\partial x + \partial E_x/\partial y = - \nabla \times E$ (4.35)

tells us that the interior circular wire (shown in grey) will receive an induced electromotive force which produces a clockwise current on it.

This is in accordance with Lenz´s law which states that the induced current tries to compensate the cause of the induction (in this case, producing a clockwise current).

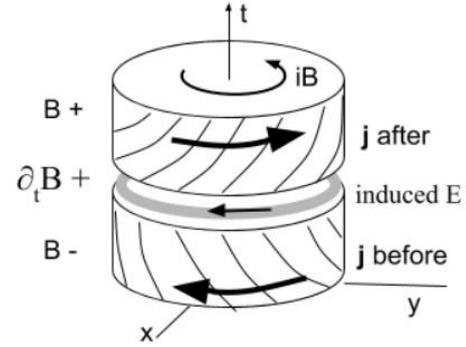

Figure 17: (1,2) spacetime as in figure 14 showing a circular current rotating clockwise which suddenly changes to a counter-clockwise rotation. The induced field E generates an electromotive force which produces a clockwise current **j**. The magnetic field due to this induced current, in accordance with Lenz´s law, tries to compensate the change in the magnetic field B due to the changing current.

## 4.5 Vector potentials in two dimensions: electromagnetic waves

The potential vector function in a two-dimensional world has the form $A = A_0 \gamma_0 + A_1 \gamma_1 + A_2 \gamma_2$ (4.36)

The corresponding field equation (Eq. 1.5), $\nabla^2 A = J$ is

$(\partial_0^2 - \partial_1^2 - \partial_2^2)(A_0\gamma_0 + A_1\gamma_1 + A_2\gamma_2) =$

$= J_0 \gamma_0 + J_1 \gamma_1 + J_2 \gamma_2$ (4.37)

Separating the different components we get three similar Poisson´s differential equations:

$(\partial_0^2 - \partial_1^2 - \partial_2^2)A_k = J_k$ ,   with k having values (0, 1, 2)

We consider again first a chargeless space (vacuum):  J = 0

In this case, all three differential equations reduce to the same form: $(\partial_0^2 - \partial_1^2 - \partial_2^2)A = 0$ or, equivalently,

$\partial^2 A/\partial t^2 = \partial^2 A/\partial x^2 + \partial^2 A/\partial y^2$ (4.38)

These are two-dimensional Laplace´s equations, known also as wave equations, whose solutions can be expressed as linear combinations of plane waves travelling at the speed of light.

Let us take one of these solutions (a general plane wave travelling to the right on the x-axis) to explore its physical meaning:

$A = A_0 \gamma_0 + A_1 \gamma_1 + A_2 \gamma_2$ (4.39)

$A = a_0(x-t)\gamma_0 + a_1(x-t)\gamma_1 + a_2(x-t)\gamma_2$ (4.40)

Applying the Lorenz gauge condition (Eq. 1.4): $\nabla \cdot A = 0$

we get $-a'_0 - a'_1 - 0 = 0$ , so that $a'_0 = -a'_1$ (4.41)

and, except for a constant factor, we can write the wave potential as $A = a_0(x-t)(\gamma_0-\gamma_1) + a_2(x-t)\gamma_2$ (4.42)

and the corresponding electromagnetic field will be

$F = \nabla \wedge A = (\partial_0\gamma_0 + \partial_1\gamma_1 + \partial_2\gamma_2) A =$

$= (\partial_0\gamma_0 + \partial_1\gamma_1) [a_0(x-t)(\gamma_0-\gamma_1) + a_2(x-t)\gamma_2] =$

$= -a'_0\gamma_{01} - a'_0\gamma_{01} - a'_2\gamma_{02} + a'_2\gamma_{12} = a'_2(-\gamma_{02} + \gamma_{12})$ (4.43)

This expression can be written in the multivector spacetime flavour as the product of a null direction $(-\gamma_0+\gamma_1)$ with a vector direction $\gamma_2$ : $F = a'_2(-\gamma_0 + \gamma_1)\gamma_2$ (4.44)

Alternatively, in the relative vector formulation we can express F as the sum of an electric field with a direction $(-\gamma_{02} = \sigma_2)$ and a magnetic field with a direction $(\gamma_{12} = \sigma_{21})$, both having the same numerical value $a'_2$ (in natural units).

We can consider a pure plane wave solution as an approximation which is valid when we are very far away from the localized source charge-current J. When this approximation is not valid, the general solution will be a superposition or linear combination of such plane waves.

A general approach to radiation should begin with the Poisson's equations which we have already seen having the form of Eq 4.37: $(\partial_0^2 - \partial_1^2 - \partial_2^2)A_k = J_k$ (4.45)

McDonald [37], citing Hadamard [38, 39] and Ehrenfest [40, 41], stated that in this 2-dimensional case the general solution has contributions from times earlier than the purely retarded time. This behaviour is clearly different from that of the radiation in 3 spatial dimensions, where only retarded times contribute to the radiation in potentials. This difference, nevertheless, vanishes when the distance from the source gets high enough, as we have seen.

The same effect applies to the case of a vanishing loop current, which in the 3D case produces a vanishing magnetic field, but in the 2D situation the magnetic field has always a non-vanishing value.

### 4.5.1 Visualization of vector potentials in two dimensions

After presenting the concept of retarded time in the one-dimensional world, where it was useful to explain the concept of redshift, it can be used in the two-dimensional case to explain further concepts like the production and characteristics of plane electromagnetic waves.

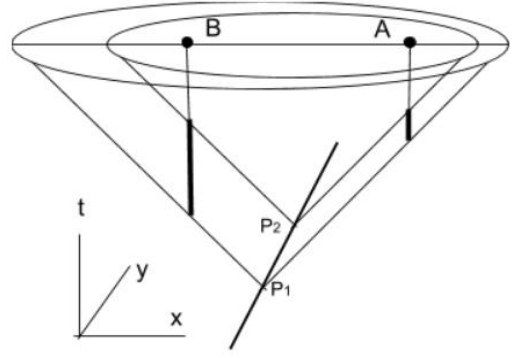

Figure 18: (1,2) spacetime as explained in figure 4 showing the potential created by a point charge P moving to the right from $P_1$ to $P_2$. Two light cones start at these points the same as in figure 4. Recall from figures 9 and 10 all the relevant information about retarded time and the Doppler effect.

Figure 18 is the equivalent of figure 10 for the two-dimensional spacetime. We can see the light cones produced by a uniformly moving charge in two instants $P_1$ and $P_2$ and how they are perceived by two observers A and B.

Figure 19 tries to present this same situation in the two-dimensional space, where the light cones are drawn as circles and the radial lines correspond to the electric field of the charged particle. We can appreciate that the distortion of the field lines is due to the displacement of the circles, which is a consequence of the movement of the source.

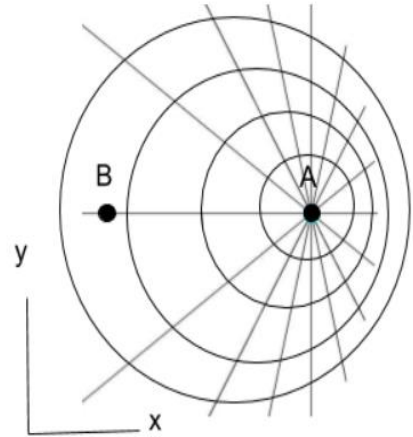

Figure 19: (2-D) space as in figure 12 (upper horizontal plane). This figure is a spatial projection of figure 18. The diminishing circles represent sections of successive light cones centred on P´s position at the moment each cone was originated. As a result, they are increasingly displaced to the right. The Doppler effect can be seen here as a change in the wavelength (increasing backwards and decreasing to the front side of the wave). The electric field lines tend to concentrate on a direction which is perpendicular to the movement of the source, increasing the strength of E.

Electromagnetic waves are produced when the source undergoes any change from the uniform movement.

These Liénard-Wiechert potentials and fields are described for the two-dimensional case by McDonald [37]. For the three-dimensional space, they were derived via Lorentz transformations [42] and discussed thoroughly in [43]. Figure 20 depicts one of the simplest changes, consisting in a brief displacement into a space direction and an equally brief returning displacement. This is called a burst.

The displacement of the light cones following this burst produces a displacement of the field lines which is maximal in the transversal direction and diminishes in intensity until it apparently disappears in the direction of the burst. This explains the absence of electromagnetic waves in the one-dimensional case.

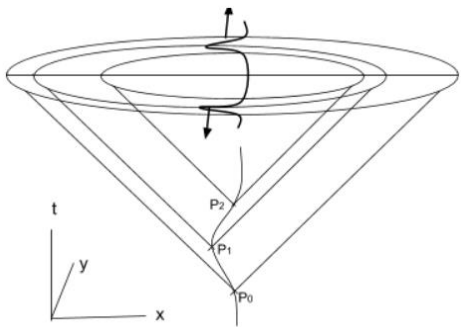

Figure 20: (1,2) spacetime as explained in figures 4 and 18. A burst is shown as a lateral displacement in the x-direction on an otherwise stationary source, beginning in $P_0$, having its maximum extension at $P_1$ and returning to the initial position at $P_2$. Stretching the corresponding light cones vertically through the time axis delivers a spatial figure (upper horizontal plane, displayed in figure 21) where the burst is seen propagating at the speed of light in both directions of the y-axis.

The same situation is presented in the two-dimensional space in figure 21, where we can see the displacement of the light cones (represented by circles, as in figure 19), and the distortions in the field lines caused by the burst.

In this figure, the burst happens in the horizontal (x) direction) direction, while the distortion is greater in the vertical direction (y) transversal to the burst, decreasing as we shift the direction to the horizontal axis, where there is no distortion at all.

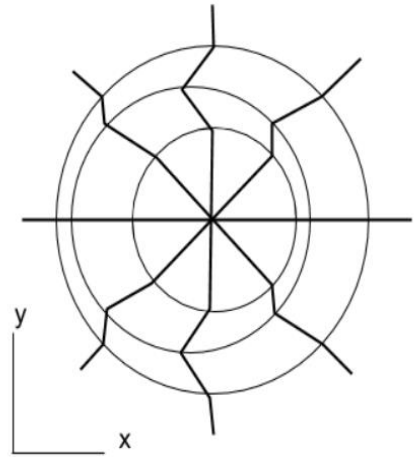

Figure 21: (2-D) space as in figure 19, showing the spatial propagation of the burst represented in spacetime on figure 20. The burst took place in the x-direction, but it does not propagate in this direction (confirming our previous one-dimensional results of section 3.3, equation 3.32). The burst propagates radially from the stationary source, with a maximum intensity in the transversal direction y.

Figure 22 depicts the result of an oscillating charged particle, which can be seen as a series of consecutive bursts. The light cones from different positions of the source, when seen in a horizontal plane, produce an oscillating pattern.

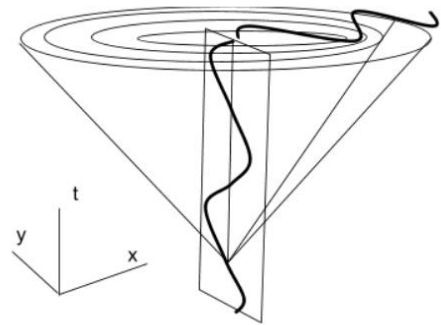

Figure 22: (1,2) spacetime as explained in figures 4, 18 and 20, representing a stationary source undergoing an oscillatory movement as a series of consecutive bursts happening, in this case, on the y-axis. Physically this corresponds to an oscillating dipole, which produces a transversal wave propagating in the x-direction, as shown in the upper horizontal space section.

When a charged particle is displaced by a forcing field, the position formerly occupied by the particle remains with an opposite charge, like a small instantaneous dipole. This is the reason for calling it a dipole oscillation.

Figure 23 shows this more clearly, where the displacement of the charge from the equilibrium position produces a dipolar field E. The field points in the same direction as the displacement because in this case we have chosen the oscillating particle, for convenience, to have a negative charge.

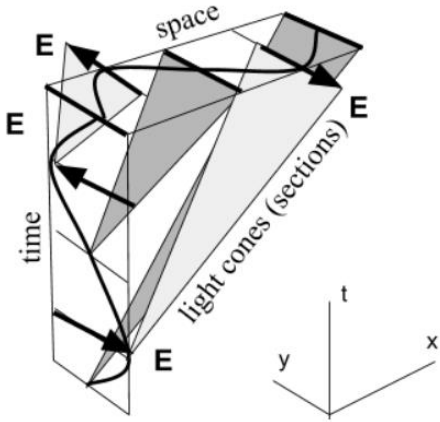

Figure 23: (1,2) spacetime as explained in figure 4 showing the same situation as in figure 22 with more detail. The lateral vertical section shows the oscillating charge as a sinusoidal curve and the electric field **E** of the resulting dipole as arrows pointing alternatively to the positive and negative directions of the y-axis. These effects propagate at the speed of light as shown by the inclined sections of the displaced light cones. The result can be seen at the upper horizontal section as an electric field oscillating in the y-direction and propagating at the speed of light through the x-axis.

We have seen that in the relative vector formulation the wave field F is the sum of an electric field **E** with a direction

$-\gamma_{02} = \sigma_2$  and a magnetic field iB with a direction $i = \sigma_{12}$

both having the same numerical value $a_2'$.

This is shown in figure 24, where a plane wave travelling in the x-axis presents a transversal oscillating electric field (straight arrows) and a corresponding magnetic field (curved arrows). The positive direction $\sigma_2$ of the electric field is shown as a vertical arrow pointing upwards at the left side, surrounded by a counter-clockwise circular arc representing the $\sigma_{12}$ direction of the magnetic field B. Both magnitudes have opposite directions at the right side of the figure.

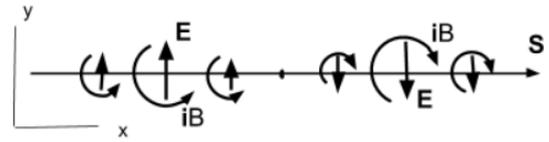

Figure 24: (2-D) space as in figure 21, showing the same situation as the one projected on the upper spatial plane of figure 23. We can recognize an electric field **E** oscillating in the y-direction and propagating at the speed of light in the positive direction of the x-axis. The field **E** is accompanied by a magnetic field B of the same magnitude (in natural units). Their directions are explained in the text. The result is an electromagnetic plane wave. Comparison with figure 15 shows that this combination of the E and B fields results in a Poynting vector signalling an energy flow in the positive x-direction. The points where both field vanish are called nodes.

Comparing with figure 15, it is possible to identify visually the propagating direction of the wave with the direction of energy transfer in an electric circuit. This is shown in the figure by the Poynting vector **S**.

A comparison of figures 23 and 24 with figure 16 shows that the electric field changes with time as a wave passes through a fixed spatial point, creating a displacement current. This leads us to understand that the points where electric and magnetic fields vanish (called nodes) are occupied by a displacement current **j**$_D$ as shown in figure 25. The displacement current **j**$_D = \partial \mathbf{E}/\partial t$ appeared already in the one-dimensional world (section 3.2, figure 8) as a chargeless current with a compensating effect of a drift current, and again in the two-dimensional world (section 4.3, figure 16) as a chargeless current which compensates the interruption of the current in a charging capacitor. In this case there is no current to be compensated for, but nevertheless the displacement current fills an apparent gap in the nodes of EM waves (where both field magnitudes E and B vanish, helping us to understand the behaviour and propagation of the EM waves in vacuum.

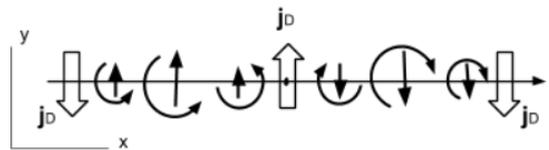

Figure 25: (2-D) space with the same situation of figure 24. The displacement current created at the nodes due to the time variation of the electric field (compare with figure 16) completes a dynamical representation of an EM plane wave, where the three magnitudes **E**, B and **J**$_D$ make alternative contributions to its existence and propagation.

# 5. Three-dimensional electromagnetism

We will use throughout this section the symbol i for the pseudoscalar $i = i_3 = \sigma_{123}$ (5.1)

Using Eq (2.14), $i = i_{1,3}$ (5.2)

In a three-dimensional world the same symbol i represents thus both pseudoscalars. As we have previously seen (Eqs 2.5, 2.12), this is not the case for other dimensions.

For the sake of simplicity, we introduce in the three-dimensional case the convention of representing the spatial indexes (1,2,3) of the vectorial components by the letter k, and the components of a given multivector direction by the ordered letters (a,b,c), grouping the various directions in a common symbol, as in

$$\gamma_a \gamma_b = \gamma_{ab} \quad (5.3)$$

$$\gamma_a \gamma_b \gamma_c = \gamma_{abc} \quad (5.4)$$

Appendix B shows a more detailed calculation using these notations; we have there also completely separated the indexes in order to check easily the resulting expressions.

The electromagnetic magnitudes that take part of the field equation in the general three-dimensional case are

$$\nabla = \partial_0 \gamma_0 + \partial_k \gamma_k$$

$$F = F_{0k} \gamma_{0k} + F_{ba} \gamma_{ba}$$

$$J = J_0 \gamma_0 + J_k \gamma_k \quad (5.5)$$

where

$$\partial_0 = \partial/\partial t, \quad \partial_k = \partial/\partial x_k \quad (5.6)$$

The space-time split gives the following correspondences:

$$J\gamma_0 = (J_0\gamma_0 + J_k\gamma_k)\gamma_0 = \rho + (-\mathbf{j}) \quad (5.7)$$

where $\mathbf{j} = j_k \sigma_k$

(Eq 1.2): $F = F_{0k}\sigma_k + F_{ba}\sigma_{ab} = \mathbf{E} + i\mathbf{B}$ (5.8)

where $\mathbf{E} = E_k\sigma_k$ and $\mathbf{B} = B_k\sigma_k$

$$\gamma_0 \nabla = \partial/\partial t + \nabla \quad (5.9)$$

Spacetime and spatial components are related by

$$F_{0k} = E_k, \quad F_{ba} = B_c, \quad J_0 = \rho, \quad J_k = -j_k \quad (5.10)$$

The field equation (Eq. 1.6) in this three-dimensional world would be: $\nabla F = J = J_0\gamma_0 + J_k \gamma_k$ (5.11)

which can be separated in four multivector directions

timelike vectors: $(\nabla \cdot F)_t = J_0 \gamma_0$

spacelike vectors: $(\nabla \cdot F)_s = J_k \gamma_k$

timelike trivectors: $(\nabla \wedge F)_t = 0$

spacelike trivectors: $(\nabla \wedge F)_s = 0$ (5.12)

## 5.1 Timelike vectors: Gauss´s law

The timelike vector component of the three-dimensional field equation can be written as:

$$(\nabla \cdot F)_t = \partial_k F_{0k} \gamma_0 = J_0 \gamma_0 \quad (5.13)$$

implying $\partial_k F_{0k} = J_0$ (5.14)

Translating to spatial components: $\partial E_k/\partial x_k = \rho$ (5.15)

or, equivalently: $\nabla \cdot \mathbf{E} = \text{div}(\mathbf{E}) = \rho$ (5.16)

which is Gauss´s law.

The physical meaning of this expression is the same as in the two-dimensional case, with the difference that the divergence operator extends now to the whole space instead of a single plane.

## 5.2 Spacelike vectors: Ampère´s law

The spacelike component of the three-dimensional field equation (Eqs. 5.12) corresponds to the spacelike components $J_k$ of the current vector J: $(\nabla \cdot F)_s = J_k \gamma_k$ (5.17)

implying $\partial_0 F_{0a} - \partial_b F_{ba} + \partial_c F_{ac} = J_a$ (5.18)

Translating to spatial components

$$\partial E_a/\partial t - \partial B_c/\partial x_b + \partial B_b/\partial x_c = -j_a \quad (5.19)$$

Rearranging terms and recalling that

$$\nabla = \sigma_k \partial/\partial x_k \quad \text{and} \quad \mathbf{B} = B_k \sigma_k \quad (5.20)$$

$$\nabla \wedge \mathbf{B} = (-\partial B_b/\partial x_a + \partial B_a/\partial x_b)\sigma_{ba} \quad (5.21)$$

$$-i(\nabla \wedge \mathbf{B}) = (\partial B_b/\partial x_a - \partial B_a/\partial x_b)\sigma_c \quad (5.22)$$

$$\partial \mathbf{E}/\partial t = \sigma_k \partial E_k/\partial t \quad (5.23)$$

$$\mathbf{j} = j_k \sigma_k \quad (5.24)$$

we can rewrite the spatial vector term of the field equation as

$$-i(\nabla \wedge \mathbf{B}) = \partial \mathbf{E}/\partial t + \mathbf{j} \quad (5.25)$$

The definition of the cross product as shown in Appendix A,

(Eq A2.19): $\mathbf{v} \times \mathbf{w} = -i(\mathbf{v} \wedge \mathbf{w})$

allows us to write $\nabla \times \mathbf{B} = -i(\nabla \wedge \mathbf{B})$ (5.26)

and get $\nabla \times \mathbf{B} = \partial \mathbf{E}/\partial t + \mathbf{j}$ (5.27)

which is Ampère´s equation.

We introduced an alternative definition for the cross-product (to allow a smooth transition between three and two dimensions) which allowed us to write

$$\nabla \times \mathbf{B} = -i(\nabla \wedge \mathbf{B}) = -\nabla \cdot (i\mathbf{B}) \quad (5.28)$$

It is interesting to compare both definitions at the same example we have already seen:

$$-\nabla \cdot (i\mathbf{B}) = -\nabla \cdot (\sigma_{abc}\mathbf{B}) = -\nabla \cdot (\sigma_{abc}B_k\sigma_k) =$$

$$= -(\sigma_k \partial/\partial x_k) \cdot (B_a\sigma_{bc}) = (\partial B_b/\partial x_a - \partial B_a/\partial x_b)\sigma_c \quad (5.29)$$

which coincides with $-i(\nabla \wedge \mathbf{B})$

### 5.3 Timelike trivectors: Faraday´s law

Any trivector part of the field equation must vanish identically, and this is true both for the timelike and the pure spatial trivectors.

The timelike trivector expression at the three-dimensional field equation (Eqs 5.12) is: $(\nabla \wedge F)_t = 0$

or, equivalently, $\partial_0 F_{ba} = -\partial_a F_{0b} + \partial_b F_{0a}$ (5.30)

Translated to the corresponding vector magnitudes:

$$\partial B_c/\partial t = -\partial E_b/\partial x_a + \partial E_a/\partial x_b \quad (5.31)$$

The right sides are the components of the cross-product with the vector derivative

$$\nabla \times \mathbf{E} = -i(\nabla \wedge \mathbf{E}) = (\partial E_b/\partial x_a - \partial E_a/\partial x_b)\sigma_c \quad (5.32)$$

we can thus write $\partial_t \mathbf{B} = -\nabla \times \mathbf{E}$ (5.33)

which is Faraday´s law of induction.

### 5.4 Spacelike trivectors: Gauss´s law for the magnetic field

The spacelike trivector expression at the three-dimensional field equation (Eqs 5.12) must also be equally null, so:

$(\nabla \wedge F)_s = 0$ (5.34)

or, equivalently, $\partial_a F_{cb} = 0$ (5.35)

Translated to the corresponding vector magnitudes:

$\partial B_k/\partial x_k = 0$ (5.36)

which can be expressed also as $\nabla \cdot \mathbf{B} = 0$ (5.37)

Which is the Gauss law for the magnetic field (implying the non-existence of magnetic sources like magnetic monopoles).

### 5.5 Vector potentials and electromagnetic waves

The potential vector function in a three-dimensional world has the form $A = A_0 \gamma_0 + A_k \gamma_k$ (5.38)

The field equation (Eq. 1.5) using this potential becomes

$\nabla^2 A = J$ (5.39)

Or $(\partial_0^2 - \partial_k^2)(A_0\gamma_0 + A_k\gamma_k) = J_0 \gamma_0 + J_k \gamma_k$ (5.40)

Separating the different components we get four Poisson´s differential equations:

$(\partial_0^2 - \partial_k^2) A_0 = J_0$ (5.41)

$(\partial_0^2 - \partial_k^2) A_k = J_k$ (5.42)

The most interesting case appears again when we consider a chargeless space (vacuum): $J = 0$

In this case, all four differential equations reduce to the same form:

$(\partial_0^2 - \partial_k^2)A = 0$ or, equivalently, $\partial^2 A/\partial x_0^2 - \partial^2 A/\partial x_k^2 = 0$

$\partial^2 A/\partial t^2 = \partial^2 A/\partial x^2 + \partial^2 A/\partial y^2 + \partial^{2a} A/\partial z^2$ (5.43)

This is the three-dimensional Laplace´s equation, whose solutions can be expressed as linear combinations of plane waves travelling at the speed of light.

The form of any of those plane waves is similar to the one we have already seen for the two-dimensional case.

## 6. Conclusions

We have tried to show how geometric algebra and geometric calculus can be used to present a smooth sequence for the understanding of the basic concepts of electromagnetism, by doing what we call a dimensional scaffolding. This was possible with the application of the synthetic formulation for the electromagnetic field equation provided by geometric algebra and geometric calculus in different dimensions.

In any dimension, we begin with the spacetime multivector formulation of the field equation, separating it into homogeneous parts and translating the results to the usual language of vector calculus.

In this process, the fundamental electromagnetic magnitudes and concepts have been appearing gradually, showing their main characteristics and allowing for their visualization.

The two-dimensional vector derivative as presented by McDonald [37] has been deduced using geometric algebra and it has proven to have deep explanatory and visual

## 6.1 Gauss´s law: charge, electric field and divergence

In one dimension, the electric field appears as a temporal bivector in spacetime with a boost-like effect, together with the concepts of spatial derivative and charge density in figure 6, where it is also possible to visualize the concept of divergence. Charges are clearly seen as sources for the electric field, and a toy model for a capacitor has been also presented in figure 7.

In two dimensions, the concept of divergence gets a clearer visualization by comparison between the spacetime diagram and its planar space-time split (figure 12).

## 6.2 Ampère´s law: magnetic field, displacement current

The concepts of temporal derivative and current density appear already in one dimension and could be explained in figure 8 as the application of a simple boost to the diagram for Gauss´s law. The same figure allows us to visualize a particular and intriguing feature of the charge-current vector in spacetime, which is reflected with respect to the worldliness of the charged particles. This is an unavoidable characteristic of spacetime, as we show in Appendices A and B, but it has no counterpart in the physical description of electromagnetism. Although in one dimension there is no equivalent to a magnetic field, the elusive concept of a chargeless displacement current could be introduced, showing a characteristic compensating attitude and taking part in different phenomena like drift currents (figure 8), charging capacitors (figure 16) and EM waves (figure 25).

In two dimensions, the magnetic field makes its appearance as a purely spatial bivector with a rotating attitude in figures 13 and 14, allowing for the introduction and basic understanding of new concepts like solenoids, capacitors and electric circuits (figures 14 to 16). The flow of energy represented by the Poynting vector, as well as the displacement current in a charging capacitor could even be addressed at this early level, showing how the magnetic field results from the combined effect of both charge and displacement currents.

## 6.3 Faraday´s law: electromotive force, induction

The magnetic field is absent in one dimension, and so does Faraday´s law, but in two dimensions it has been possible to explain -with help of a spacetime diagram as in figure 14 - the induction of an electromotive force as a consequence of a time-varying magnetic field. The resulting induced current stands against the change that produces it, and this is in accordance with Lenz´s law too.

## 6.4 Poisson´s equation: potentials, waves, retarded time, Doppler effect.

The potential formulation for the field equation in one dimension allowed us to arrive in this extremely simple situation to the concept of wave equation, whose solutions are plane waves with the speed of light as shown in figure 9. Retarded time could also be presented and visualized in figure 10, allowing for the understanding of the redshift and retarded potentials. No physical meaning, nevertheless, could be assigned to those potential waves in a one-dimensional world because they render identically null electric fields.

In two dimensions, the potentials for moving charges could be visualized as effects propagating with the speed of light in figures 18 to 22. The analysis of a plane wave solution together with its physical meaning as a combination of equal-valued electric and magnetic fields normal to the propagating direction were shown in figures 23 to 25.

## 6.5 Dimensional scaffolding sequence

In order to visualize the general idea that has guided this work, figure 26 represents the scaffold that we have built along the precedent lines. The whole sequence that has been followed is drawn as a three-dimensional diagram whose meaning is explained in the caption.

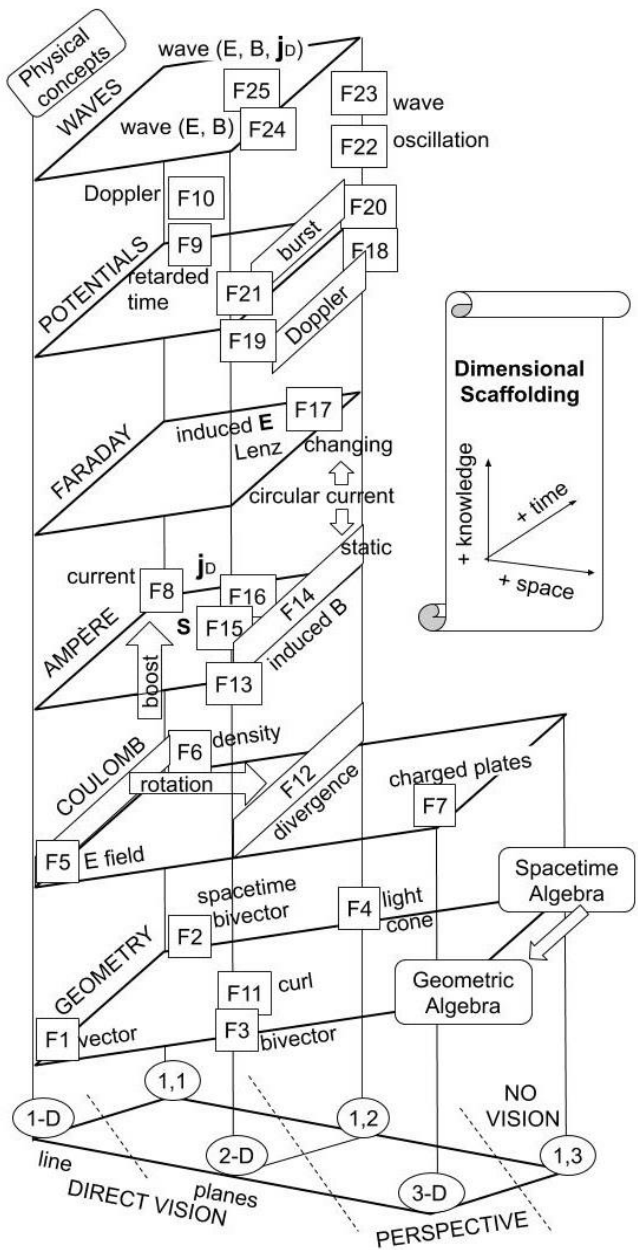

Figure 26: Visual representation of the dimensional scaffolding process in a purely illustrative diagram without physical or geometric meaning. The squares with an F followed by a number represent the figures as numbered throughout the paper. They are arranged horizontally according to their space/time dimensionality (labelled in circles at the bottom plane), and vertically according to their physical content (concepts, laws, magnitudes). The bottom plane is divided by dashed lines into four sections: Two (lines and planes) for the figures that don´t need any special strategy to be understood, another one for the three-dimensional figures, which need the convention of perspective to be created and understood, and the last one for the four-dimensional (1,3) diagrams corresponding to the physical world as explained by Minkowski. They are absent in this paper because our mind is not prepared to manage four dimensions.

## 6.6 Final remarks

We have shown how the gradual presentation of the fundamental electromagnetic concepts by a scaffolding process over increasing dimensions has been possible in a unified way using the extremely simplicity of the geometric algebra formulations for the field and potential equations. Nevertheless, the whole algebraic three-dimensional structure should be needed to introduce and explain some remaining concepts like the complementarity between electric and magnetic fields –as well as the absence of magnetic monopoles-, electromagnets and electric motors, rotating wires and generators, and wave polarization, for example.

## 7. Acknowledgements

We thank Kirk Mc Donald, from the Department of Physics of Princeton University, for fruitful discussions and the incorporation of one of our views as additional remarks in [37], and Rosana Rodríguez López, from the Departamento de Estatística, Análise Matemática e Optimización of the Universidade de Santiago de Compostela, for her help.

## References

[1] Grassmann H 1844 *Die Lineale Ausdehnungslehre ein neuer Zweig der Mathematik Dargestellt und durch Anwendungen auf die übrigen Zweige der Mathematik, wie auch auf die Statik, Mechanik, die Lehre vom Magnetismus und die Krystallonomie erläutert* (Leipzig: Wiegand)

[2] Grassmann H 1862 *Die Ausdehnungslehre: Vollständig und in Strenger Form Bearbeitet* (Berlin: Enslin)

[3] Clifford W K 1878 *American Journal of Mathematics* **1** (4) 350-358

[4] Clifford W K On the Classification of Geometric Algebra *Mathematical Papers by W. K. Clifford* ed R Tucker (London) p 397-401

[5] Hamilton W R 1844 On Quaternions; or on a new System of Imaginaries in Algebra. Letter from Sir William R. Hamilton to John T. Graves *The London, Edinburgh and Dublin Philosophical Magazine and Journal of Science* vol **xxv** p 489-495

[6] Hamilton W R 1844-1850 On Quaternions; or on a new System of Imaginaries in Algebra, *The London, Edinburgh and Dublin Philosophical Magazine and Journal of Science* (it appeared in 18 instalments in volumes xxv-xxxvi of *The London, Edinburgh and Dublin Philosophical Magazine and Journal of Science* (3rd Series), for the years 1844-1850).

[7] Hamilton W R 1853 *Lectures on Quaternions* (Dublin: Hodges and Smith)

[8] Francis, M R and Kosowsky A 2005 The construction of spinors in geometric algebra. *Annals of Physics* **317** 383

doi: https://doi.org/10.1016/j.aop.2004.11.008

[9] Chappell J M, Iqbal A, Hartnett J G and Abbott D 2016 The Vector Algebra War: A Historical Perspective *IEEE Access* **4** 1997-2004

doi: 10.1109/ACCESS.2016.2538262.

[10] Gibbs J W 1881-1884 Elements of Vector Analysis: Arranged for the Use of Students in Physics (Privately printed, New Haven: Tuttle, Morehouse & Taylor) numbers 101, 1881; 102-189 and note

[11] Heaviside O 1893 *Electromagnetic Theory* (London) vol I

[12] Hestenes D, Sobczyk G 1984 *Clifford Algebra to Geometric Calculus, a Unified Language for Mathematics and Physics* (The Netherlands: Springer)

[13] Hestenes D 1986 *A Unified Language for Mathematics and Physics*. In *Clifford Algebras and their Applications in Mathematical Physics* ed J S R Chisholm and A K Common (Dordrecht/Boston: Reidel) 1-23

[14] Aste A 2012 Complex representation theory of the electromagnetic field *Journal of Geometry and Symmetry in Physics* **28**: 47–58

arXiv:1211.1218. doi:10.7546/jgsp-28-2012-47-58.

[15] Belkovich I V and Kogan B L 2016 Utilization of Riemann-Silberstein Vectors in Electromagnetics *Progress In Electromagnetics Research B* **69** 103–116

[16] Bialynicki-Birula I and Bialynicka-Birula Z 2013 The role of the Riemann-Silberstein vector in classical and quantum theories of electromagnetism *Journal of Physics A Mathematical and Theoretical* **46** 053001

doi: 10.1088/1751-8113/46/5/053001

[17] Doran C and Lasenby A 2007 Geometric Algebra for Physicists. Cambridge University Press. ISBN 10: 0521480221

doi: https://doi.org/10.1017/CBO9780511807497

[18] Macdonald A 2017 A Survey of Geometric Algebra and Geometric Calculus. *Adv. Appl. Clifford Algebras* **27** 853–891

doi: https://doi.org/10.1007/s00006-016-0665-y

[19] Hestenes D *Primer on Geometric Algebra* http://geocalc.clas.asu.edu/pdf/PrimerGeometricAlgebra.pdf

[20] Hestenes D 1968 *J. Math. Anal. and Appl.* **24** (2) 313–325

[21] Hestenes D 2002 *Oersted Medal Lecture 2002: reforming the Mathematical Language of Physics* http://geocalc.clas.asu.edu/pdf/OerstedMedalLecture.pdf

[22] Wheeler N *"Electrodynamics" in 2-dimensional spacetime* https://www.reed.edu/physics/faculty/wheeler/documents/Electrodynamics/Miscellaneous%20Essays/E&M%20in%202%20Dimensions.pdf

[23] Penrose R 2004 The Road to Reality: A Complete Guide to the Laws of the Universe. Alfred A. Knopf Publisher, New York. ISBN: 978-0679454434

[24] Mureika J and Stojkovic D 2011 Detecting Vanishing Dimensions via Primordial Gravitational Wave Astronomy. *Phys. Rev. Lett*. **106** 101101

doi:https://doi.org/10.1103/PhysRevLett.106.101101

[25] Stojkovic D 2013 Vanishing Dimensions: A Review. *Modern Physics Letters A* 28(37) 1330034

doi: https://doi.org/10.1142/S0217732313300346

[26] Minkowski, H. 2012. *Space and Time. Minkowskí´s Papers on Relativity* (free version). Translated by Fritz Lewertoff and Vesselin Petkov. Edited by Vesselin Petkov. Canada: Minkowski Institute Press.

[27] Naber G 2012 *The geometry of Minkowski spacetime. An introduction to the mathematics of the special theory of relativity*. 2nd ed. Berlin: Springer.

[28] Hestenes D 1992 *Mathematical Viruses* In: *Clifford Algebras and their Applications in Mathematical Physics* (The Netherlands: Kluwer Academic Publishers) 3-16. DOI: 10.1007/978-94-015-8090-8_1

[29] Prado X, Area I, Paredes A, Domínguez J M, Edelstein J and Mira J 2018 Archimedes meets Einstein: a millennial geometric bridge. *Eur. J. Phys.* **39** 045802

doi: https://doi.org/10.1088/1361-6404/aab12c

[30] Constantinescu F 1980 *Distributions and their Applications in Physics.* Edited by Farina J E G and Fullerton G H. Pergamon

doi: https://doi.org/10.1016/C2013-0-02647-3

[31] Grubb G 2009 *Distributions and Operators,* Springer.

[32] Jackson J D and Okun L B 2001 Historical roots of gauge invariance *Reviews of Modern Physics* **73**(3) 663–680 doi:10.1103/RevModPhys.73.663

[33] Aharonov Y and Bohm D 1959 Significance of electromagnetic potentials in quantum theory *Physical Review* **115**(3) 485–491

doi: https://doi.org/10.1103/PhysRev.115.485

[34] Batelaan H and Tonomura A 2009 The Aharonov–Bohm effects: Variations on a subtle theme. *Physics Today* **62**(9) 38

doi: https://doi.org/10.1063/1.3226854

[35] Pearle P and Rizzi A 2017 Quantized vector potential and alternative views of the magnetic Aharonov-Bohm phase shift. *Phys. Rev. A* **95** 052124

doi: https://doi.org/10.1103/PhysRevA.95.052124

[36] Wheeler N. 1996 *Electrodynamical Applications of the Exterior Calculus.*

https://www.reed.edu/physics/faculty/wheeler/documents/Electrodynamics/Miscellaneous%20Essays/Exterior%20E&M.pdf

[37] McDonald K *Electrodynamics in 1 and 2 spatial dimensions*

http://kirkmcd.princeton.edu/examples/2dem.pdf

[38] Hadamard J 1923 *Lectures on Cauchy's Problem in Linear Partial Differential Equations* (Yale U. Press)

[39] Hadamard J 1908 Theorie des ́Equations aux Dérivées Partielles Linéaires Hyperboliques et du Probleme de Cauchy *Acta Math.* **31** 333

[40] Ehrenfest P 1918 In that way does it become manifest in the fundamental laws of physics that space has three dimensions? *Proc. Kön. Neder. Akad. Wet.* **20** 200

[41] Ehrenfest P 1920 Welche Rolle spielt die Dreidimensionalität des Raumes in den Grundgesetzen der Physik? *Ann. Phys.* **61** 440

[42] 2017 McDonald K *Liénard-Wiechert Potentials and Fields via Lorentz Transformations*

http://www.hep.princeton.edu/~mcdonald/examples/lw_potentials.pdf

[43] McDonald K *The Relation Between Expressions for Time-Dependent Electromagnetic Fields Given by Jefimenko and by Panofsky and Phillips*

http://kirkmcd.princeton.edu/examples/jefimenko.pdf

[44] Jackson J D 1998. *Classical Electrodynamics* Third Edition, Wiley

[45] Kitano M 2012 Reformulation of Electromagnetism with Differential Forms (Trends in Electromagnetism - From Fundamentals to Applications).

https://www.intechopen.com/books/trends-in-electromagnetism-from-fundamentals-to-applications/reformulation-of-electromagnetism-with-differential-forms

[46] Sattinger D H 2013 Maxwell's Equations, Hodge Theory, and Gravitation. Semantic Scholar. https://api.semanticscholar.org/CorpusID:119298776. https://arxiv.org/pdf/1305.6874.pdf

[47] Landau L D and Lifshitz E 1987 The Classical Theory of Fields. Elsevier, Amsterdam. ISBN: 9780750627689

[48] Hjemeland S E and Lindstrom U Duality for non-specialists. hep-th/9705122

https://ui.adsabs.harvard.edu/abs/1997hep.th....5122H/abstract

[49] Hehl F and Obukhov Y 2003 Foundations of Classical Electrodynamics: Charge, Flux and Metric. Springer Science+Business Media, LLC

doi: 10.1007/978-1-4612-0051-2

[50] Baylis W 1998 Electrodynamics. A Modern Geometric Approach. Springer, Birkhäuser Basel. ISBN 978-0-8176-4025-5

[51] Sha W, Wu X L, Huang Z X and Chen M S 2008 Maxwell's Equations, Symplectic Matrix, and Grid. Progress *Electromagnetics Research B* **8** 115–127

doi:10.2528/PIERB08052303

[52] Kachkachi M 1997 Symplectic structure of electromagnetic duality. https://arxiv.org/abs/hep-th/9707140v1

ç

# Appendix A

## Main features of geometric algebra and notations used in this work

Geometric algebra is based mainly on the concept of multivector as well as the existence of the geometric product and its strong property of being associative.

The geometric product of two vectors can be decomposed as

$$ab = a \cdot b + a \wedge b \tag{A.1}$$

where $a \cdot b = b \cdot a$ is the inner product, whose result is a scalar.

and $a \wedge b = - b \wedge a$ is the outer product, whose result, as we will see, is called a bivector.

The elements of an n-dimensional geometric algebra are called multivectors, and they can be expressed in a basis of $2^n$ elements, which can be arranged in blades of the same dimension such as scalars (dim = 0), vectors (dim =1), bivectors (dim = 2), trivectors (dim =3) or, alternatively, by decreasing dimensionalities such as pseudoscalars (dim = n), pseudovectors (dim = n-1), and so on.

This is because the dimension of every blade is the number of vector directions that must be combined by the geometric product to produce the respective directions.

For example, if the space has dimension N=3, the possible blades are scalars (dim=0), vectors (dim =1), bivectors. (dim=2) and trivectors (dim =3).

The respective combinations of indexes are: none for the unique scalar direction, (1,2,3) for the three possible vector directions, (12, 23, 31) for the three bivector directions, and (123) for the unique trivector direction (which is also a pseudoscalar in this 3-dimensional space).

The sum of all these numbers of multivector directions gives $1+3+3+1 = 8 = 2^3$.

The number of directions of a D-dimensional blade D in a space of dimension N is the number of possible combinations of N elements taken in groups of D. The ordered arrangement of these dimensions gives the Pascal´s Triangle, where every horizontal line corresponds to a certain space dimension N, and the elements are its blades in increasing order of their dimension D.

The sum of all the elements in a horizontal line gives the dimension of the multivector algebra in the corresponding N-dimensional space:

$2^0, 2^1, 2^2, 2^3, 2^4$, etc.

```
        1
       1 1
      1 2 1
     1 3 3 1
    1 4 6 4 1
```
Pascal`s Triangle

The elements of any of these blades can be expressed in an orthonormal basis, beginning with scalars (whose basis consists only of the number 1) and vectors (whose basis is a set of n orthonormal vectors.

We followed throughout the text the convention of applying the symbol $\gamma$ for the spacetime vector directions and the boldface symbol $\boldsymbol{\sigma}$ for the relative vector directions of its associated even subalgebra. Readers who are familiar with the geometric algebra of the Euclidean three-dimensional space should be careful in order to avoid confusions.

In a space of n dimensions, the spacetime basis consists of (n+1) elements $\{\gamma_0, \gamma_1, \gamma_2, ...\gamma_n\})$. These basis elements or vector directions have the following properties:

$$\gamma_j \cdot \gamma_k = 0 \tag{A.2}$$

$$\gamma_i \gamma_i = \gamma_i \cdot \gamma_i = \pm 1 \tag{A.3}$$

(+1 for i = 0 and -1 for i = 1,2,3, …) This fundamental distinction between a timelike direction $\gamma_0$ and a spacelike direction $\gamma_{i \neq 0}$ is a characteristic feature of spacetime. An important consequence of it is the existence of null-directions like, for example, $\gamma_0 + \gamma_1$.

$(\gamma_0+\gamma_1)^2 = (\gamma_0+\gamma_1)(\gamma_0+\gamma_1) = \gamma_0\gamma_0+\gamma_0\gamma_1+\gamma_1\gamma_0+\gamma_1\gamma_1=$

$$= 1+0+0-1 = 0 \tag{A.4}$$

This is the reason to call them "null directions", and they form the so-called "null-cone" or "light cone".

Elements of higher dimensionality are consistently created by means of the outer product: $\gamma_k \gamma_l = \gamma_k \wedge \gamma_l = -\gamma_l \gamma_k$   (A.5)

In this process, when two equal indexes are adjacent, they are removed and the direction is multiplied by the sign of their product. Two different adjacent indexes can exchange positions, changing the sign of the direction.

Bivectors have basis elements like

$\{\gamma_1\gamma_0, \gamma_2\gamma_0, \gamma_3\gamma_0, \gamma_2\gamma_1, \gamma_3\gamma_2, \gamma_1\gamma_3\}$.

The inner product, in general, is the grade-lowering part of the geometric product, and the outer product is its grade-rising part, as we have seen for vectors. This is even true for a vector (v) and a bivector (W), whose geometric product can be decomposed as

$$vW = v \cdot W + v \wedge W \quad (A.6)$$

where $v \cdot W = -W \cdot v$ is the inner product and $v \wedge W = W \wedge v$ is the outer product.

The commutation properties, therefore, are not the same as in the case of vectors.

This is due to the properties of the direction products.

In the case of vectors, $\gamma_a \gamma_b = \gamma_{ab} = -\gamma_{ba} = -\gamma_b \gamma_a$ (A.7)

for different indexes, while $\gamma_a \gamma_a = \gamma_a \gamma_a$ for equal indexes is evident.

In the case of a vector with a bivector,

$\gamma_a \gamma_{bc} = \gamma_{abc} = \gamma_{bca} = \gamma_{bc} \gamma_a$ for different indexes, (A.8)

whilst $\gamma_a \gamma_{ab} = \gamma_{aab} = -\gamma_{aba} = \gamma_{ab} \gamma_a$ (A.9)

and $\gamma_a \gamma_{ba} = \gamma_{aba} = -\gamma_{baa} = \gamma_{ba} \gamma_a$ (A.10)

if one of the indexes is shared.

A1. Geometric calculus

Geometric calculus is founded on the concept of a vector derivative $\nabla$, which can be expressed as

$$\nabla = \partial_0 \gamma_0 + \partial_1 \gamma_1 + \partial_2 \gamma_2 \ldots \quad (A1.1)$$

where $\partial_i$ means the partial derivative with respect to the coordinate $x_i$

$$\partial_i = \partial/\partial x_i \quad (A1.2)$$

In the case of an orthonormal basis on a flat manifold, this is the ordinary gradient $\partial_k \gamma_k$ together with the time derivative $\partial_0 \gamma_0$.

The operator $\nabla$ can be treated as a vector:

$$\nabla^2 = \nabla \cdot \nabla = \partial_0^2 - \partial_1^2 - \partial_2^2 - \partial_3^2 =$$
$$= \partial^2/\partial x_0^2 - \partial^2/\partial x_1^2 - \partial^2/\partial x_2^2 - \partial^2/\partial x_3^2 \quad (A1.3)$$

where the signs arise from the products $\gamma_i \gamma_i$.

The Fundamental Theorem of Geometric Calculus, which embodies in a common way all sort of Stoke´s-type formulas in different dimensions and signatures, opens the possibility of inverting the vectorial derivative. The resulting antiderivative depends mainly on the use of the appropriate Green´s functions.

A2. Space-time split

In order to connect spacetime with the usual framework of an euclidean space together with a separate time coordinate, we will use the so-called space-time split. This begins with the selection of a certain timelike direction in spacetime which we will call for convenience $\gamma_0$. It can be identified with the observer's reference system and it is sometimes called simply "the observer".

The split is made by the geometric product of the observer with the physical magnitudes in spacetime. To see how this works we will present a pair of examples, the first for a spacetime vector V and the second for a spacetime bivector F.

The split of a vector V is produced by

$$V\gamma_0 = V \cdot \gamma_0 + V \wedge \gamma_0 = v_0 + \mathbf{v}_k \quad (A2.1)$$

where $v_0$ is a scalar and $\mathbf{v}_k$ is a vector in the euclidean space. These vectors are written in boldface type, and they are called "relative vectors" to distinguish them from the previously seen spacetime vectors. The reason is that they depend on the - relative - choice of the observer $\gamma_0$.

We introduce the notation $i_n$ to identify specifically the pseudoscalar in the n-dimensional euclidean space and $i_{1,n}$ for the (1,n)-dimensional spacetime pseudoscalar, the former being the even subalgebra of the latter.

If $\{\gamma_0, \gamma_1, \ldots, \gamma_n\}$ form an orthonormal basis for the (1,n)-dimensional spacetime, then its pseudoscalar will be given by the expression

$$i_{1,n} = \gamma_0 \gamma_1 \ldots \gamma_n = \gamma_0 \wedge \gamma_1 \wedge \ldots \wedge \gamma_n \quad (A2.2)$$

We can build a basis for the corresponding n-dimensional euclidean space defining its elements as

$$\boldsymbol{\sigma}_k = \gamma_{k0} = \gamma_k \gamma_0 = \gamma_k \wedge \gamma_0 \quad (A2.3)$$

In the three-dimensional space [8], we write these relative vector directions as ($\boldsymbol{\sigma}_1, \boldsymbol{\sigma}_2, \boldsymbol{\sigma}_3$)

The orthogonality is assured because of

$$\boldsymbol{\sigma}_k^2 = \boldsymbol{\sigma}_k \boldsymbol{\sigma}_k = \gamma_k \gamma_0 \gamma_k \gamma_0 = -\gamma_k \gamma_k \gamma_0 \gamma_0 = 1 \quad (A2.4)$$

(using $\gamma_0 \gamma_0 = 1$ and $\gamma_k \gamma_k = -1$)

$$\boldsymbol{\sigma}_k \boldsymbol{\sigma}_l = \gamma_k \gamma_0 \gamma_l \gamma_0 = -\gamma_k \gamma_l \gamma_0 \gamma_0 = -\gamma_k \gamma_l = \gamma_l \gamma_k \quad (A2.5)$$

$$\boldsymbol{\sigma}_l \boldsymbol{\sigma}_k = \gamma_l \gamma_0 \gamma_k \gamma_0 = -\gamma_l \gamma_0 \gamma_0 \gamma_k = -\gamma_l \gamma_k = -\boldsymbol{\sigma}_k \boldsymbol{\sigma}_l \quad (A2.6)$$

We can now define the pseudoscalar

$$i_n = \boldsymbol{\sigma}_{1\ldots n} = \boldsymbol{\sigma}_1 \ldots \boldsymbol{\sigma}_n = \boldsymbol{\sigma}_1 \wedge \ldots \wedge \boldsymbol{\sigma}_n \quad (A2.7)$$

Both types of pseudoscalar are related:

$$i_1 = \boldsymbol{\sigma}_1 = \gamma_{10} = -\gamma_{01} = -i_{1,1} \quad (A2.8)$$

$i_2 = \sigma_{12} = \gamma_{1020} = -\gamma_{0012} = -\gamma_0 \, i_{1,2}$ (A2.9)

$i_3 = \sigma_{123} = \gamma_{102030} = i_{1,3}$ (A2.10)

The euclidean space is 3-dimensional, therefore we can use $i_3$ or $i_{1,3}$ indistinctly and write i – somehow abusing the notation for sake of simplicity - to represent both cases, but for lower dimensionalities this is not the case, and we should specify which type of pseudoscalar ($i_n$ or $i_{1,n}$) is being used.

Following Hestenes [13] (p. 14), the space-time split of F is

$F = \mathbf{E} + i\mathbf{B}$ (A2.11)

Where **E** anticommutes with $\gamma_0$ and ($i\mathbf{B}$) commutes with it:

$\mathbf{E}\gamma_0 = -\gamma_0 \mathbf{E}$, $\quad i\mathbf{B}\,\gamma_0 = \gamma_0\, i\mathbf{B}$ (A2.12)

$\gamma_0 F \gamma_0 = \gamma_0(\mathbf{E} + i\mathbf{B})\gamma_0 = \gamma_0 \mathbf{E}\gamma_0 + \gamma_0(i\mathbf{B})\gamma_0 = -\mathbf{E}\gamma_0\gamma_0 + (i\mathbf{B})\gamma_0\gamma_0$

$\gamma_0 F \gamma_0 = -\mathbf{E} + i\mathbf{B}$ (A2.13)

This allows us to write

$\mathbf{E} = \tfrac{1}{2}(F - \gamma_0 F \gamma_0)$ and $i\mathbf{B} = \tfrac{1}{2}(F + \gamma_0 F \gamma_0)$ (A2.14)

$\mathbf{E} = E\sigma_k$ is a relative vector with direction $\sigma_k = \gamma_{k0} = -\gamma_{0k}$

whereas $i\mathbf{B} = B\sigma_{ab}$ is a relative bivector with direction

$\sigma_{ab} = \gamma_{a0b0} = -\gamma_{ab00} = -\gamma_{ab}$ (A2.15)

Multiplying $i\mathbf{B}$ by the pseudoscalar $i = i_n$ allows us to get the magnetic field component B, which has a different nature in every dimensional scaffolding step:

-For the one-dimensional case, there is no possible spatial bivector $\gamma_{ab}$ and thus no magnetic field at all.

-For the two-dimensional case, $i\mathbf{B} = B\sigma_{12}$ and the pseudoscalar $i = i_2 = \sigma_{12}$ ($i^2 = -1$) (A2.16)

$\mathbf{B} = -i^2\mathbf{B} = -i(i\mathbf{B}) = -\sigma_{12}(B\,\sigma_{12}) = -\sigma_{12}^2\, B = B$, so that the magnetic field component is the scalar magnitude B

-For the three-dimensional case, $i\mathbf{B} = B_1\sigma_{23} + B_2\sigma_{31} + B_3\sigma_{12}$ and the pseudoscalar $i = i_3 = \sigma_{123}$ ($i^2 = -1$) (A2.17)

$\mathbf{B} = -i^2\mathbf{B} = -i(i\mathbf{B}) = -\sigma_{123}(B_1\sigma_{23} + B_2\sigma_{31} + B_3\sigma_{12}) = B_1\sigma_1 + B_2\sigma_2 + B_3\sigma_3$ so that the magnetic field component is a relative vector.

The space-time split of the vector derivative $\nabla$ yields:

$\nabla\gamma_0 = \nabla\cdot\gamma_0 + \nabla\wedge\gamma_0 = \partial_o + \partial_k\gamma_{k0} =$

$= \partial/\partial t + \sigma_1 \partial/\partial x + \sigma_2 \partial/\partial y + \sigma_3 \partial/\partial z = \partial/\partial t + \boldsymbol{\nabla}$ (A2.18)

The cross-product of two relative vectors can be defined as

$\mathbf{v}\times\mathbf{w} = -i(\mathbf{v}\wedge\mathbf{w})$ (A2.19)

The following identity: $\mathbf{v}\wedge\mathbf{w} = -(i\mathbf{v})\cdot(i\mathbf{w})$ (A2.20)

allows us to use an alternative but equivalent definition for the cross product: $\mathbf{v}\times\mathbf{w} = -\mathbf{v}\cdot(i\mathbf{w})$ (A2.21)

This alternative definition will be useful for the building of an analogue to the cross-product in the two-dimensional case.

A3. Electromagnetism in geometric algebra

The set of Maxwell´s Equations can be unified in the following expression: $\nabla F = J$

This is the electromagnetic field equation in geometric algebra, and in this context we can call it simply the field equation.

The symbol $\nabla$ stands for the spacetime vector derivative as presented into the framework of geometric calculus.

F is a bivector field: $F = F_{0k}\gamma_{0k} + F_{ba}\gamma_{ba}$ (A3.1)

which should be interpreted like a sum of terms where *k* can have the values (1,2,3) and we define the pairs of subindexes *ba* as having the values (21, 32, 13)

With help of a space-time split, this formula can be translated to relative vectors, getting in this way the four Maxwell equations in the usual vector formulation.

Natural units (like Heaviside-Lorentz) where

$\mu_0 = 1$ and $\varepsilon_0 = 1$

are being used consistently and, consequently, the speed of light $c = 1$.

Applying a space-time split to the field bivector F (Eq A2.11) yields $F = \mathbf{E} + i\mathbf{B}$

where **E** is a relative vector (the electric field), corresponding to the temporal component of the bivector F, and **B** is the magnetic field vector. The product $i\mathbf{B}$ corresponds to the purely spatial part of the spacetime bivector F.

The symbol i corresponds to the euclidean pseudoscalar ($i_n$ in our notation), in order to keep the expression valid for dimensions other than 3.

F is thus splitted in two parts, the first part corresponding to the electric field **E**: $F_{0k} = E_k$ (A3.2)

and the second part to the magnetic field **B**:

$F_{ba} = B_c$ or, equivalently,

$F_{21} = B_3, \quad F_{32} = B_1, \quad F_{13} = B_2$ (A3.3)

J is a vector field: $J = J_0\gamma_0 + J_i\gamma_i$ (A3.4)

$J\gamma_0 = J \cdot \gamma_0 + J \wedge \gamma_0 = \rho + (-\mathbf{j})$ (A3.5)

where the scalar part corresponds to the charge density ($J_0 = \rho$), and the relative vector part to the vector current density $\mathbf{j}$: $J_i = -j_i$ (A3.6)

The negative sign appearing at this last identity is needed to assure a perfect correspondence between the spacetime formulation and the usual relative vectorial set of Maxwell´s formulas.

The bivector F can be derived from a vector-valued potential function A by means of the formula: $\nabla A = F$ (A3.7)

together with a gauge to eliminate ambiguities.

We will follow the Lorenz gauge: $\nabla \cdot A = F$ (A3.8)

The form of the field equation in GA using this vector potential is $\nabla F = \nabla(\nabla A) = \nabla^2 A = J$ (A3.9)

## A4. Correspondence between spacetime and purely spatial magnitudes

After establishing the EM field equations in a n-dimensional world using its (1,n) dimensional spacetime algebra, we restore its associated (n-dimensional) even subalgebra (by space-time splitting) in order to present the results in the language of relative (Gibbs) vectors. In each step we don´t change the dimensionality (n) of the real world being considered, but the dimensionality of the GA we use for it: (1,n) for spacetime algebra, (n) for Gibbs relative vector algebra (even subalgebra).

We begin with the electromagnetic equations on both formulations: $\nabla F = J$ (A4.1)

where

$\nabla = \partial_0 \gamma_0 + \partial_1 \gamma_1 + \partial_2 \gamma_2 + \partial_3 \gamma_3 =$

$= \partial_0 \gamma_0 + \partial_k \gamma_k$  (k = 1,2,3)  (A4.2)

$F = F_{01} \gamma_{01} + F_{02} \gamma_{02} + F_{03} \gamma_{03} + F_{21} \gamma_{21} + F_{32} \gamma_{32} + F_{13} \gamma_{13} =$

$= F_{0k} \gamma_{0k} + F_{ba} \gamma_{ba}$ (A4.3)

(k =1,2,3) and (ab =21, 32, 13)

$J = J_0 \gamma_0 + J_1 \gamma_1 + J_2 \gamma_2 + J_3 \gamma_3 =$

$= J_0 \gamma_0 + J_k \gamma_k$   (k = 1,2,3)   (A4.4)

for the spacetime multivector formulation, and the set of four Maxwell´s laws for the relative vector formulation:

Gauss: $\nabla \cdot \mathbf{E} = \rho$ (A4.5)

Faraday: $\nabla \times \mathbf{E} = -\partial \mathbf{B}/\partial t$ (A4.6)

Ampère: $\nabla \times \mathbf{B} = \mathbf{j} + \partial \mathbf{E}/\partial t$ (A4.7)

Gauss for B: $\nabla \cdot \mathbf{B} = 0$ (A4.8)

where

$\nabla = \sigma_1 \partial/\partial x + \sigma_2 \partial/\partial y + \sigma_3 \partial/\partial z = \sigma_k \partial_k$  is the relative vector derivative

$\mathbf{E} = E_1 \sigma_1 + E_2 \sigma_2 + E_3 \sigma_3 = E_k \sigma_k$ (A4.9)

$\mathbf{B} = B_1 \sigma_1 + B_2 \sigma_2 + B_3 \sigma_3 = B_k \sigma_k$ (A4.10)

are the electric and magnetic fields, and $\times$ is the usual cross-product, which can be expressed also as $\mathbf{v} \times \mathbf{w} = -i\mathbf{v} \wedge \mathbf{w}$, being i the pseudoscalar $i = \sigma_1 \sigma_2 \sigma_3$ (A4.11)

Beginning with Gauss´s law for the electric field: $\nabla \cdot \mathbf{E} = \rho$

$\sigma_k \partial_k E_k \sigma_k = \partial_k E_k = \rho$ (A4.12)

the corresponding term in spacetime is the timelike vector relation:

$\partial_k \gamma_k F_{0k} \gamma_{0k} = J_0 \gamma_0$ (A4.13)

$\partial_k F_{0k} \gamma_k \gamma_{0k} = -\partial_k F_{0k} \gamma_{kk0} = \partial_k F_{0k} \gamma_0$ (A4.14)

and $\partial_k F_{0k} = J_0$ (A4.15)

identifying $J_0$ with the charge density $\rho$, it follows that

$F_{0k} = E_k$ (A4.16)

($F_{01} = E_1$ , $F_{02} = E_2$ , $F_{03} = E_3$ ) (A4.17)

Faraday´s equation $\nabla \times \mathbf{E} = -\partial \mathbf{B}/\partial t$ can be written as

$\partial_a E_b - \partial_b E_a = \partial_t B_c$  where the indexes (a, b, c) stand for (1, 2, 3), (2, 3, 1) or (3, 1, 2) in order to write down three different terms.

In spacetime, the corresponding equation has to be obtained from the timelike trivector part:

$\partial_0 F_{ba} \gamma_{0ba} + \partial_a F_{0b} \gamma_{a0b} + \partial_b F_{0a} \gamma_{b0a} = 0$ (A4.18)

(because J is only a vector-valued function in spacetime, without trivector parts)

Rearranging terms to get the same direction:

$\partial_0 F_{ba} \gamma_{ba0} + \partial_a F_{0b} \gamma_{ba0} - \partial_b F_{0a} \gamma_{ba0} = 0$ (A4.19)

we get $\partial_0 F_{ba} + \partial_a F_{0b} - \partial_b F_{0a} = 0$ (A4.20)

where we can substitute $F_{0k}$ with $E_k$ to obtain

$\partial_0 F_{ba} + \partial_a E_b - \partial_b E_a = 0$ (A4.21)

and $\partial_a E_b - \partial_b E_a = -\partial_0 F_{ba}$ (A4.22)

A direct comparison with the relative vectorial expression obtained earlier

$\partial_a E_b - \partial_b E_a = \partial_t B_c$ (A4.23)

allows us to write $F_{ba} = B_c$ (A4.24)

($F_{21} = B_3$ , $F_{32} = B_1$ , $F_{13} = B_2$)

Recalling finally Ampère's law

$\nabla \times \mathbf{B} = \mathbf{j} + \partial \mathbf{E}/\partial t$ (A4.25)

which can be written as

$\partial_a B_b - \partial_b B_a = j_c + \partial_t E_c$ (A4.26)

using the same convention for the combinations of subindexes, we can compare it with the purely spatial vector expression in spacetime

$\partial_0 F_{0c}\gamma_{00c} + \partial_b F_{cb}\gamma_{bcb} + \partial_a F_{ac}\gamma_{aac} = (\partial_0 F_{0c} + \partial_b F_{cb} - \partial_a F_{ac})\gamma_c =$

$= J_c \gamma_c$ (A4.27)

$\partial_0 F_{0c} + \partial_b F_{cb} - \partial_a F_{ac} = J_c$ (A4.28)

$\partial_a F_{ac} - \partial_b F_{cb} = - J_c + \partial_0 F_{0c}$ (A4.29)

we can substitute again the spacetime components with their space equivalents as obtained previously to get

$\partial_a B_b - \partial_b B_a = - J_c + \partial_0 E_c$ (A4.30)

and, recalling our previous expression $\partial_a B_b - \partial_b B_a = j_c + \partial_t E_c$

we arrive to the equivalence

$J_c = - j_c$ (A4.31)

($J_1 = - j_1$ , $J_2 = - j_2$ , $J_3 = - j_3$) (A4.32)

We have already introduced this minus sign in section 3, where we explained that it does not imply any change in the physical magnitude **j**. Its meaning can be interpreted as the spacetime vector J being spatially reflected with respect to the current vector, and this is a striking geometric feature with no physical counterpart.

## Appendix B

## Detailed calculations using components

### B1. Ordering the spacetime geometric algebra directions

As we have seen, geometric algebra offers the possibility of expressing the formulas of electromagnetism (EM) in very synthetic coordinate-free expressions which hold formally for several dimensions.

Nevertheless, more detailed calculations using coordinates are also possible, and we include here some calculations of the results we have seen previously as an exercise for readers who are not yet familiar with the subtleties of geometric algebra and want to check where they arise from and how they function.

The following tables may be helpful to understand in a straightforward way how the directions combine in geometric algebra, applied to the field equation in the four-dimensional spacetime and its corresponding three-dimensional space.

The field equation $\nabla F = J$ bears a geometric product between the vector derivative $\nabla$

$\nabla = \partial_0\gamma_0 + \partial_1\gamma_1 + \partial_2\gamma_2 + \partial_3\gamma_3$

and the field bivector F

$F = F_{01}\gamma_{01}+F_{02}\gamma_{02}+F_{03}\gamma_{03}+F_{23}\gamma_{23}+F_{31}\gamma_{31}+F_{12}\gamma_{12}$

The result of a geometric product between a vector and a bivector consists, in general, in the sum of a vector v and a trivector T:

$\nabla F = v+T$

Where $v = \nabla \cdot F$, $T = \nabla \wedge F$

In the special case of the electromagnetic field equation, the vector part is the charge-current vector J, and the trivector part is equally null: $\nabla F = \nabla \cdot F = J$

A series of tables will show us how the four vector directions ($\gamma_0$, $\gamma_1$, $\gamma_2$, $\gamma_3$) combine with the six bivector directions

($\gamma_{01}$, $\gamma_{02}$, $\gamma_{03}$, $\gamma_{23}$, $\gamma_{31}$, $\gamma_{12}$).

The first step consists in creating all the possible ordered arrays of the indexes, as shown in table 1:

|  | $\gamma_{01}$ | $\gamma_{02}$ | $\gamma_{03}$ | $\gamma_{23}$ | $\gamma_{31}$ | $\gamma_{12}$ |
|---|---|---|---|---|---|---|
| $\gamma_0$ | $\gamma_{001}$ | $\gamma_{002}$ | $\gamma_{003}$ | $\gamma_{023}$ | $\gamma_{031}$ | $\gamma_{012}$ |
| $\gamma_1$ | $\gamma_{101}$ | $\gamma_{102}$ | $\gamma_{103}$ | $\gamma_{123}$ | $\gamma_{131}$ | $\gamma_{112}$ |
| $\gamma_2$ | $\gamma_{201}$ | $\gamma_{202}$ | $\gamma_{203}$ | $\gamma_{223}$ | $\gamma_{231}$ | $\gamma_{212}$ |
| $\gamma_3$ | $\gamma_{301}$ | $\gamma_{302}$ | $\gamma_{303}$ | $\gamma_{323}$ | $\gamma_{331}$ | $\gamma_{312}$ |

Table 1: geometric product of vector and bivector directions.

We apply now the rules for the combination and exchange of indexes in order to write them in a coherent way in table 2:

$\gamma_{00} = 1$, $\gamma_{kk} = -1$ (for k = 1, 2, 3),

$\gamma_{rs} = -\gamma_{sr}$ (for r ≠ s)

|  | $\gamma_{01}$ | $\gamma_{02}$ | $\gamma_{03}$ | $\gamma_{23}$ | $\gamma_{31}$ | $\gamma_{12}$ |
|---|---|---|---|---|---|---|
| $\gamma_0$ | $\gamma_1$ | $\gamma_2$ | $\gamma_3$ | $\gamma_{023}$ | $\gamma_{031}$ | $\gamma_{012}$ |
| $\gamma_1$ | $\gamma_0$ | $-\gamma_{012}$ | $\gamma_{031}$ | $\gamma_{123}$ | $\gamma_3$ | $-\gamma_2$ |
| $\gamma_2$ | $\gamma_{012}$ | $\gamma_0$ | $-\gamma_{023}$ | $-\gamma_3$ | $\gamma_{123}$ | $\gamma_1$ |
| $\gamma_3$ | $-\gamma_{031}$ | $\gamma_{023}$ | $\gamma_0$ | $\gamma_2$ | $-\gamma_1$ | $\gamma_{123}$ |

Table 2: combination and reordering of indexes

We can simplify this into table 3 by taking into account the properties of the pseudoscalar $i = i_3 = i_{1,3} = \gamma_{0123}$

$i\gamma_0 = \gamma_{01230} = -\gamma_{123}$ , $i\gamma_1 = \gamma_{01231} = -\gamma_{023}$ ,

$i\gamma_2 = \gamma_{01232} = -\gamma_{031}$ , $i\gamma_3 = \gamma_{01233} = -\gamma_{012}$

|  | $\gamma_{01}$ | $\gamma_{02}$ | $\gamma_{03}$ | $\gamma_{23}$ | $\gamma_{31}$ | $\gamma_{12}$ |
|---|---|---|---|---|---|---|
| $\gamma_0$ | $\gamma_1$ | $\gamma_2$ | $\gamma_3$ | $i\gamma_1$ | $i\gamma_2$ | $i\gamma_3$ |
| $\gamma_1$ | $\gamma_0$ | $-i\gamma_3$ | $i\gamma_2$ | $-i\gamma_0$ | $\gamma_3$ | $-\gamma_2$ |
| $\gamma_2$ | $i\gamma_3$ | $\gamma_0$ | $-i\gamma_1$ | $-\gamma_3$ | $-i\gamma_0$ | $\gamma_1$ |
| $\gamma_3$ | $-i\gamma_2$ | $i\gamma_1$ | $\gamma_0$ | $\gamma_2$ | $-\gamma_1$ | $-i\gamma_0$ |

Table 3: introducing the pseudoscalar i

We can recognize in table 3 the following groups of directions:

Timelike vectors $<\nabla\cdot F>_t$: direction $\gamma_0$

Spacelike vectors $<\nabla\cdot F>_s$: directions $(\gamma_1, \gamma_2, \gamma_3)$

Spacelike trivectors $<\nabla\wedge F>_s$: direction $i\gamma_0 = -\gamma_{123}$

Mixed trivectors $<\nabla\wedge F>_t$: directions $(i\gamma_1, i\gamma_2, i\gamma_3)$

Now we write the partial derivatives of the bivector components in table 4 following the same order.

|  | $F_{01}$ | $F_{02}$ | $F_{03}$ | $F_{23}$ | $F_{31}$ | $F_{12}$ |
|---|---|---|---|---|---|---|
| $\partial_0$ | $\partial_0 F_{01}$ | $\partial_0 F_{02}$ | $\partial_0 F_{03}$ | $\partial_0 F_{23}$ | $\partial_0 F_{31}$ | $\partial_0 F_{12}$ |
| $\partial_1$ | $\partial_1 F_{01}$ | $\partial_1 F_{02}$ | $\partial_1 F_{03}$ | $\partial_1 F_{23}$ | $\partial_1 F_{31}$ | $\partial_1 F_{12}$ |
| $\partial_2$ | $\partial_2 F_{01}$ | $\partial_2 F_{02}$ | $\partial_2 F_{03}$ | $\partial_2 F_{23}$ | $\partial_2 F_{31}$ | $\partial_2 F_{12}$ |
| $\partial_3$ | $\partial_3 F_{01}$ | $\partial_3 F_{02}$ | $\partial_3 F_{03}$ | $\partial_3 F_{23}$ | $\partial_3 F_{31}$ | $\partial_3 F_{12}$ |

Table 4: partial derivatives of the field bivector F

## B2. Comparison of spacetime and spatial expressions

The final expressions will be obtained by the combination of the terms in table 4 that share the same group of indexes in table 3.

The resulting vector $J = \nabla\cdot F$ has two parts:

-A timelike component $<\nabla\cdot F>_t$ with direction $\gamma_0$, $(\partial_1 F_{01} + \partial_2 F_{02} + \partial_3 F_{03})\gamma_0 = J_0 \gamma_0$

which can be written as $\partial_k F_{0k} = J_0$ (F1)

-A spacelike component $<\nabla\cdot F>_s$ whose direction is a combination of $(\gamma_1, \gamma_2, \gamma_3)$

$(\partial_0 F_{01} + \partial_2 F_{12} - \partial_3 F_{31}) \gamma_1 = J_1 \gamma_1$

$(\partial_0 F_{02} + \partial_3 F_{23} - \partial_1 F_{12}) \gamma_2 = J_2 \gamma_2$

$(\partial_0 F_{03} + \partial_1 F_{31} - \partial_2 F_{23}) \gamma_3 = J_3 \gamma_3$

Which can be written as

$\partial_0 F_{0a} + \partial_b F_{ab} - \partial_c F_{ca} = J_a$ (F2)

Where the indexes (a,b,c) can have the following values: (1,2,3), (2,3,1) or (3,1,2)

The directions showing the pseudoscalar i correspond to the components of the trivector parts, whose combinations are identically null for the electromagnetic field equation.

The corresponding equation $\nabla\wedge F = 0$ can be separated also in two parts:

-A purely spacelike component $<\nabla\wedge F>_s$ with direction

$i\gamma_0 = -\gamma_{123}$

$(\partial_1 F_{23} + \partial_2 F_{31} + \partial_3 F_{12}) i\gamma_0 = 0$

which can be written as $\partial_a F_{ab} = 0$ (F3)

-A partially timelike component $<\nabla\wedge F>_t$ whose direction is a combination of

$(i\gamma_1, i\gamma_2, i\gamma_3)$

$(\partial_0 F_{23} + \partial_3 F_{02} - \partial_2 F_{03}) i\gamma_1 = 0$

$(\partial_0 F_{31} + \partial_1 F_{03} - \partial_3 F_{01}) i\gamma_2 = 0$

$(\partial_0 F_{12} + \partial_2 F_{01} - \partial_1 F_{02}) i\gamma_3 = 0$

Which can be written as

$\partial_0 F_{bc} + \partial_c F_{0b} - \partial_b F_{0c} = 0$ (F4)

The resulting set of field equations (F):

(F1): $\partial_k F_{0k} = J_0$

(F2): $\partial_0 F_{0a} + \partial_b F_{ab} - \partial_c F_{ca} = J_a$

(F3): $\partial_a F_{ab} = 0$

(F4): $\partial_0 F_{bc} + \partial_c F_{0b} - \partial_b F_{0c} = 0$

can be compared with the corresponding Maxwell´s equations (M):

-Gauss: $\nabla \cdot \mathbf{E} = \rho$

or, equivalently, $\partial_k E_k = \rho$ (M1)

-Ampère: $\nabla \times \mathbf{B} = \partial \mathbf{E}/\partial t + \mathbf{j}$

implying $\partial_b B_c - \partial_c B_b = \partial E_a/\partial t + j_a$ (M2)

-Gauss for B: $\nabla \cdot \mathbf{B} = 0$, or $\partial_a B_a = 0$ (M3)

-Faraday: $\nabla \times \mathbf{E} = -\partial \mathbf{B}/\partial t$

$\partial_b E_c - \partial_c E_b = -\partial B_a/\partial t$ (M4)

We can now make a stepwise comparison between the expressions resulting from the field equation (F) and those obtained from Maxwell´s laws (M) in order to establish direct coherent correspondences between the spacetime magnitudes and their related spatial equivalents.

Beginning with F1 and M1 (Gauss):

$\partial_k F_{0k} = J_0$ , $\partial_k E_k = \rho$

It is evident that the natural choice of $J_0$ as equivalent to the charge density ($J_0 = \rho$) leads to the identification between the temporal components of the field bivector and those of the electric field ($F_{0k} = E_k$).

We can now resort to F4

$\partial_0 F_{bc} + \partial_c F_{0b} - \partial_b F_{0c} = 0$, rewrite it as

$\partial_0 F_{bc} + \partial_c E_b - \partial_b E_c = 0$

or $\partial_b E_c - \partial_c E_b = \partial_0 F_{bc}$

and make a comparison with M4 (Faraday)

$\partial_b E_c - \partial_c E_b = -\partial B_a/\partial t$

The partial spacetime derivative $\partial_0$ is equivalent to the time derivative $\partial/\partial t$, and as a result the components of the magnetic field are related to the purely spatial components of the field bivector by the expression

$F_{bc} = -B_a$ which embodies three identities:

$F_{23} = -B_1$ , $F_{31} = -B_2$ , $F_{12} = -B_3$

Finally, F2: $\partial_0 F_{0a} + \partial_b F_{ab} - \partial_c F_{ca} = J_a$

rewritten as $\partial_0 E_a - \partial_b B_c + \partial_c B_b = J_a$

or $\partial_b B_c - \partial_c B_b = \partial_0 E_a - J_a$

can be compared with M2 (Ampère)

$\partial_b B_c - \partial_c B_b = \partial E_a/\partial t + j_a$

arriving finally to the expression $J_a = -j_a$

A final remark: we could have arrived to a different set of equivalences changing the signs in all of them, beginning with $J_0 = -\rho$ which applying Gauss´s law would lead to

$F_{0k} = -E_k$

Faraday´s law would require in this case that $F_{bc} = -B_a$

And finally applying Ampère´s law we would arrive to
$J_a = j_a$

Although this can seem more natural, changing the sign in $J_0$ makes that the current vector J points backwards in time, without erasing the reflection of the vector J with respect to the current lines in spacetime as shown in figure 8. This geometric feature does not depend therefore on the choice of signs.

### B3. Comparing (E,B) tables in different dimensions

Maxwell´s equations (M) can be rewritten placing every partial derivative of the EM fields at the left side:

M1 (Gauss): $\partial_k E_k = \rho$

M2 (Ampère): $\partial E_a/\partial t - \partial_b B_c + \partial_c B_b = j_a$

M3 (Gauss for B): $\partial_a B_a = 0$

M4 (Faraday): $\partial B_a/\partial t + \partial_b E_c - \partial_c E_b = 0$

We can then rewrite table 4 using the correspondence we have established between the spacetime field bivector and their relative vector EM counterparts.

The numbers inside table 5 reflect the number we assigned to each of the four Maxwell ´s equations (M), with the sign each partial derivative has in them.

|  | $E_1$ | $E_2$ | $E_3$ | $B_1$ | $B_2$ | $B_3$ |
|---|---|---|---|---|---|---|
| $\partial_t$ | 2 | 2 | 2 | 4 | 4 | 4 |
| $\partial_1$ | 1 | 4 | -4 | 3 | -2 | 2 |
| $\partial_2$ | -4 | 1 | 4 | 2 | 3 | -2 |
| $\partial_3$ | 4 | -4 | 1 | -2 | 2 | 3 |

Table 5: partial derivatives of the electric and magnetic field and Maxwell´s equations

It is possible to observe a kind of symmetric correspondence between the behaviour of the magnitudes E and B in both parts of table 5.

Formulations of EM in the language of p-forms [44, 45] have shown that there is an intrinsic geometric structure in the three-dimensional case. This is related to the Hodge dualities between the EM field magnitudes [46, 47, 48, 49, 50]. We can visualize the effect of these dualities on the (E, B) fields if we identify them with the closely related (D,H) magnitudes. Table 6 combines field quantities and differential operators as four-dimensional tensors arranging them as second-order, antisymmetric tensors $F^{\alpha\beta}$ and its Hodge duals (right part).

| 0 | $E_1$ | $E_2$ | $E_3$ | 0 | $-B_1$ | $-B_2$ | $-B_3$ |
|---|---|---|---|---|---|---|---|
| $-E_1$ | 0 | $-B_3$ | $B_2$ | $B_1$ | 0 | $E_3$ | $-E_2$ |
| $-E_2$ | $B_3$ | 0 | $-B_1$ | $B_2$ | $-E_3$ | 0 | $E_1$ |
| $-E_3$ | $-B_2$ | $B_1$ | 0 | $B_3$ | $E_2$ | $-E_1$ | 0 |

Table 6: Arrangement of the three-dimensional EM field components as antisymmetric tensors (left) and their Hodge duals (right).

We can observe a symmetric structure with resembles the one shown in table 5.

If we want to produce a table which is similar to table 6 for the two-dimensional case, we would found something like table 7. In this case, the blank square at the right side means that no similar duality can be constructed in this case, because the magnetic field has only one component.

| 0 | $E_1$ | $E_2$ |
|---|---|---|
| $-E_1$ | 0 | $-B$ |
| $-E_2$ | $B$ | 0 |

Table 7: Arrangement of the two-dimensional EM field components as an antisymmetric tensor without a clear Hodge dual.

Table 8 repeats the same visual argument for the one-dimensional EM, with no magnetic component at all.

| 0 | $E$ |
|---|---|
| $-E$ | 0 |

Table 8: Arrangement of the one-dimensional electric field as an antisymmetric tensor without a clear Hodge dual.

In the three-dimensional case the EM field magnitudes show a symplectic symmetry [51, 52] when arranged in a 2x3 file, such as ($E_1$, $E_2$, $E_3$, $B_1$, $B_2$, $B_3$).

Since symplectic symmetry affects 2n dimensional magnitudes, it is not clear how such a symmetry could be found for a similar arrangement of the EM field magnitudes in the two-dimensional ($E_1$, $E_2$, B) or one-dimensional (E) cases.